\documentclass[aps,pra,reprint,showpacs,superscriptaddress,nofootinbib,twocolumn]{revtex4-2}
\usepackage{amsmath}
\usepackage{amsfonts}
\usepackage{amsthm}
\usepackage{amssymb}
\usepackage{graphicx}
\usepackage{enumerate}
\usepackage{color}
\usepackage{lineno}
\usepackage[T1]{fontenc}
\usepackage{mathptmx}
\usepackage{braket}
\usepackage{todonotes}
\usepackage{soul}
\usepackage{subfigure}
\graphicspath{{figures/}}

\newcommand{\mysection}[1]{\textit{#1}.---}

\usepackage{float}
\usepackage{mathrsfs,amsfonts,dsfont}
\usepackage{epsfig}
\usepackage{bm}
\usepackage{framed}
\usepackage{tcolorbox}
\usepackage{multirow}
\usepackage{verbatim}
\usepackage{physics}
\usepackage[10pt]{moresize}
\usepackage{amscd}
\usepackage{amssymb}
\usepackage{tikz}
\usepackage[ruled,vlined]{algorithm2e}
\usepackage[keys,primitives]{cryptocode}
\usepackage{tabularx}
\usepackage{diagbox}
\usepackage{lipsum}
\usepackage[utf8]{inputenc}
\usepackage[bookmarks=false,colorlinks,citecolor=blue,
linkcolor=blue,anchorcolor=blue,urlcolor=blue
]{hyperref}
\begin{document}

%\title{Challenging Local Realism via Loophole-free Hardy's Violation}

\title{Loophole-free test of local realism via Hardy's violation}
%Hardy 还是Hardy's

%\date{\today}
%\author{xxx}
\author{Si-Ran Zhao}
\affiliation{Hefei National Research Center for Physical Sciences at the Microscale and School of Physical Sciences, University of Science and Technology of China, Hefei, 230026, China}
\affiliation{CAS Center for Excellence in Quantum Information and Quantum Physics, University of Science and Technology of China, Hefei, 230026, China}
\author{Shuai Zhao}
\affiliation{School of Cyberspace, Hangzhou Dianzi University, Hangzhou 310018, China}

\author{Hai-Hao Dong}
\author{Wen-Zhao Liu}
\affiliation{Hefei National Research Center for Physical Sciences at the Microscale and School of Physical Sciences, University of Science and Technology of China, Hefei, 230026, China}
\affiliation{CAS Center for Excellence in Quantum Information and Quantum Physics, University of Science and Technology of China, Hefei, 230026, China}
\author{Jing-Ling Chen}
%\email{chenjl@nankai.edu.cn}
\affiliation{Theoretical Physics Division, Chern Institute of Mathematics, Nankai University, Tianjin 300071, China}

\author{Kai Chen}
%\email{kaichen@ustc.edu.cn}
\affiliation{Hefei National Research Center for Physical Sciences at the Microscale and School of Physical Sciences, University of Science and Technology of China, Hefei, 230026, China}
\affiliation{CAS Center for Excellence in Quantum Information and Quantum Physics, University of Science and Technology of China, Hefei, 230026, China}
\affiliation{Hefei National Laboratory, University of Science and Technology of China, Hefei, 230026, China}

\author{Qiang Zhang}
%\email{qiangzh@ustc.edu.cn}
\affiliation{Hefei National Research Center for Physical Sciences at the Microscale and School of Physical Sciences, University of Science and Technology of China, Hefei, 230026, China}
\affiliation{CAS Center for Excellence in Quantum Information and Quantum Physics, University of Science and Technology of China, Hefei, 230026, China}
\affiliation{Hefei National Laboratory, University of Science and Technology of China, Hefei, 230026, China}

\author{Jian-Wei Pan}
%\email{pan@ustc.edu.cn}
\affiliation{Hefei National Research Center for Physical Sciences at the Microscale and School of Physical Sciences, University of Science and Technology of China, Hefei, 230026, China}
\affiliation{CAS Center for Excellence in Quantum Information and Quantum Physics, University of Science and Technology of China, Hefei, 230026, China}
\affiliation{Hefei National Laboratory, University of Science and Technology of China, Hefei, 230026, China}

\begin{abstract}
Bell's theorem states that quantum mechanical description on physical quantity cannot be fully explained by local realistic theories, and lays solid basis for various quantum information applications. Hardy's paradox is celebrated to be the simplest form of Bell's theorem 
concerning its "All versus Nothing" way to test local realism. However, due to experimental imperfections, existing tests of Hardy's paradox require additional assumptions of experimental systems, which constitute potential loopholes for faithfully testing local realistic theories.  Here, we experimentally demonstrate Hardy’s nonlocality through a photonic entanglement source.  By achieving a detection efficiency of 82.2\%, a quantum state fidelity of 99.10\% and applying high speed quantum random number generators for measurement setting switching, the experiment is implemented in a loophole-free manner.
%we experimentally demonstrate Hardy's nonlocality without additional assumptions in a loophole-free manner with system detection efficiency being around 82.2\%, fidelity being 99.10\% and high speed quantum random number generators. 
During $6$ hours of running,  a strong violation of $P_{\text{Hardy}}=4.646\times 10^{-4}$ up to $5$ standard deviations is observed with $4.32\times 10^{9}$ trials. A null hypothesis test shows that the results can be explained by local realistic theories with an upper bound probability of $10^{-16348}$. These testing results present affirmative evidence against local realism, and provide an advancing benchmark for quantum information applications based on Hardy's paradox.   
\end{abstract}
%%%%%%%%%%%%%%%%%%%%%%%%%%%%%%%

%\keywords{Bell nonlocality, Hardy's paradox, Quantum theory}
%%%%%%%%%%%%%%%%%%

%%%%%%%%%%%%%%%%%%
\maketitle
%\linenumbers

%\section{Introduction}
The advent of Quantum Mechanics has exerted a profound impact on our understanding of world. While, it is so counter-intuitive that there exist severe controversies for quantum theory, such as Einstein-Podolosky-Rosen's (EPR's) argument on the completeness of quantum description on physical reality~\cite{Einstein1935}. At the heart of EPR's argument is the paradox between the probabilistic description by quantum theory and the deterministic description by classical theory on physical reality. To explain the probabilistic behavior of quantum theory from classical perspective, Bohm proposed the local hidden variable (LHV) model~\cite{Bohm1952A}. Later, Bell presented an inequality as a test whether the quantum behavior can be explained by the LHV model~\cite{bell1964einstein,clauser1969proposed}. For quantum theory, the violation of a Bell inequality can be achieved, which indicates that the results from quantum theory cannot be fully explained by the LHV model. This phenomenon is known as Bell nonlocality~\cite{brunner2014bell}. The experimental demonstrations of the Bell nonloclality along the routine of Bell inequalities have been conducted soon after its derivation~\cite{Freedman1972Experimental,Aspect1982Experimental}, and recently have been pushed into the regime of loophole-free realization~\cite{hensen2015loophole,Shalm2015Strong,Giustina2015Significant,Li2018Test,Rauch2018Cosmic,storz2023loophole}, which promote a vast range of \emph{device-independent (DI) applications}~\cite{mayers1998quantum,Barrett2005no,Acin2006from,Acin2007device,colbeck2007quantum,pironio2010random,colbeck2012free,bierhorst2018experimentally,liu2018device,Zhang2020Experimental,shalm2021device,Li2021Experimental,Liu2022Toward,Xu2022Device,nadlinger2022experimental,zhang2022device,li2023device,zapatero2023advances,vsupic2023quantum}.

% From the LHV model, the joint probability description on the behavior involving two parties, Alice and Bob, which has the equivalent form $P(ab|xy)=\int d\lambda q(\lambda)P(a|x,\lambda)P(b|y,\lambda)$, 
% where $x$ and $y$ are measurement inputs, $a$ and $b$ are measurement outputs for Alice and Bob, respectively~\cite{brunner2014bell}. The $\lambda$ is a local hidden variable, which is essential for the assumption of locality. The $q(\lambda)$ is the probability distribution of $\lambda$ with $\int d\lambda q(\lambda)=1$. It is evident that the above behavior is essentially deterministic~\cite{Fine1982Hidden,Goh2018Geometry}.
% Comparatively, quantum behavior can be expressed as $P(ab|A_iB_j)=Tr(\rho_{AB}{A}_{a|i}\otimes{B}_{b|j})$,
% where $\rho_{AB}$ is quantum state for Alice and Bob's systems, ${A}_{a|i}$ (${B}_{b|j}$) is the projector for Alice's (Bob's) measurement outcome $a$ ($b$) with input $x={A}_i$ ($y={B}_j$). 

Besides Bell inequalities, there exist other approaches to demonstrate nonlocality, i.e. Bell's theorems without inequality which are emerged since the Greenberger-Horne-Zeilinger (GHZ)'s theorem~\cite{greenberger1989going,greenberger1990bell,brunner2014bell}. The GHZ's theorem pioneers an "All versus Nothing (AVN)" way to test local realism. While, at the very beginning, it's only applicable to three or more-party quantum systems. Soon after, Hardy's paradox is proposed as a "simplest version of Bell's theorem" by simultaneously keeping the "AVN" feature and being applicable to two-party systems~\cite{hardy1993nonlocality,mermin1994quantum}. Specifically, Hardy's paradox is interpreted as that the conditions
$P(00|A_2B_2)=0$, $P(01|A_1B_2)=0$, $P(10|A_2B_1)=0$, which must lead to $P(00|A_1B_1)=0$ for LHV models. While, it can maximally achieve Hardy's value $P(00|A_1B_1)=\frac{5\sqrt{5}-11}{2}$ by quantum theory~\cite{hardy1993nonlocality}. Here, $P(ab|xy)$ is the joint probability involving two parties, Alice and Bob, with $x\in\{A_1, A_2\}$ and $y\in\{B_1, B_2\}$ being measurement inputs and $a, b\in\{0, 1\}$ being measurement outputs for Alice and Bob, respectively~\cite{brunner2014bell}. Along with this fundamental interest, Hardy's paradox also finds its applications in quantum information processing including DI dimension witness, DI quantum randomness certification, DI quantum key distribution, self-testing of quantum systems and so on~\cite{Mukherjee2015Hardy,Li2015Device,Rahaman2015device,ramanathan2018practical,Rai2022self,zhao2023tilted}.

Despite great efforts have been made by experimentalists \cite{BoschiLadder1997,BARBIERI200523towards,Vallone2011Testing,Chen2012Quantum,Karimi2014Hardy,Chen2017Experimental,Yang2019Stronger,das2020new}, loophole-free Hardy's paradox test is still missing, which limits significantly its related quantum information applications. In this letter, we challenge the local realism with Hardy's violation by utilizing polarization-entangled photon pairs with a high-fidelity of up to $99.10\%$, fast random basis choices, and high detection efficiency of around $82.2\%$ to obtain the joint probabilities of Alice and Bob.
Specifically, by simultaneously closing the \emph{locality loophole} and \emph{detection loophole} and using high speed quantum random number generators (QRNGs) to guarantee random measurement setting choices,  we demonstrate a Hardy's violation of $P_{\text{Hardy}}=4.646\times 10^{-4}$ up to more than 5 standard deviations with a set of events containing $4.32\times 10^9$ trials during 6 hours running time (For local realistic theory, the Hardy's value should be $P_{\text{Hardy}}\leq 0)$. By the null hypothesis test following the prediction-based-radio (PBR) method \cite{Zhang2011Asymptotically}, the upper bound of the probability that local realistic theories can reproduce the observed Hardy's correlation is $p\leq 10^{-16348}$. These results provide strong evidence that the quantum mechanical predictions cannot be described by local realistic theories. Meanwhile, it serves as a benchmark for quantum information applications based on Hardy's paradoxes.

One of the main obstacles for loophole-free Hardy's paradox test is that, compared with the Bell inequality tests,  the theoretical analysis remains incomplete. In practice, due to imperfect detection efficiency $\eta< 1$,  there are undetected events, denoted as $u$. 
If one discards these undetected events, it will result in a severe \emph{detection loophole}. Here, to test local realism with Hardy's paradox without \emph{detection loophole}, we take these undetected events $u$ into account. During each trial, Alice and Bob choose one of two measurement settings, respectively. The measurement results for Alice and Bob are denoted as ternary elements $a, b\in\{0, 1, u\}$ respectively. Inspired by References~\cite{Eberhard1993Background,HWANG1996The}, when Hardy's conditions $P(00|A_2B_2)=0$, $P(01|A_1B_2)=0$ and $P(10|A_2B_1)=0$ are satisfied, there must be Hardy's value $P_{\text{Hardy}}=P(00|A_1B_1)-P(0u|A_1B_2)-P(u0|A_2B_1)\leq 0$,
 for local hidden variable models. While, it can achieve positive Hardy's values for quantum theory with 
 \begin{equation}\begin{split}
      P_{\text{Hardy}}^{\text{max}}(\eta)=& \frac{1}{2}[1-\sqrt{1+4\eta(3\eta-2)}]\\
                                          &+3\eta[1-3\eta+\sqrt{1+4\eta(3\eta-2)}],
 \end{split}
 \end{equation}

 for $\eta\in(2/3,1]$ (See Appendix.~\ref{Optimal_det_effi_and_state} for details).

Besides the imperfect detection efficiency, due to the dark counts of detectors and multiple pairs of photons from the entangled source, there are also some double clicks events in practical experiment, which correspond to the case where both the result 1 and 0 are obtained simultaneously in a single laboratory. Here, in order to close the \emph{detection loophole}, we designate these double clicks events as inconclusive events $u$, too. Moreover, due to such imperfections, zero Hardy's conditions are experimentally unattainable. Consequently, a form of Hardy's inequality $P_{Hardy}\leq0$ together with three Hardy’s conditions are necessary for loophole-free Hardy's paradox test. Inspired by the strategy of tackling non-zero Hardy's conditions in References~\cite{Ghirardi2006Hardy,Braun2008Hardy}, when Hardy's conditions $P(00|A_2B_2)=\epsilon_1$, $P(01|A_1B_2)=\epsilon_2$ and $P(10|A_2B_1)=\epsilon_3$ are satisfied, there must be Hardy's value
\begin{equation}\label{inequ}
\begin{split}
   P_{\text{Hardy}}=&P(00|A_1B_1)-P(0u|A_1B_2)-P(u0|A_2B_1)-\sum_{i=1}^{3}\epsilon_i\\
   &\leq 0,
\end{split}
\end{equation} for LHV models. Here, $u$ denotes inconclusive events including undetected events and double-click events, $\epsilon_i$ with $i\in\{1,2,3\}$ are small values for non-zero Hardy's conditions. 

Another obstacle is that, compared with the Bell inequality tests, the loophole-free Hardy's paradox test requires even higher detection efficiency and higher fidelity of entanglement states. For example, with the system detection efficiency ($\eta \approx 78.8\%$) and quantum state fidelity ($F\approx 98.66\%$) of Reference~\cite{Li2018Test}, we show that the Hardy's value should be less than $10^{-6}$, which is quite hard to be realized in a loophole-free manner (See Appendix.~\ref{Optimal_det_effi_and_state} for detail). In the following sections, we experimentally demonstrate a loophole-free Hardy's inequality violation by quantum mechanics statistics as a paradox against local realistic theories.

\begin{figure*}[htbp]
\centering
\includegraphics[width =0.7\textwidth]{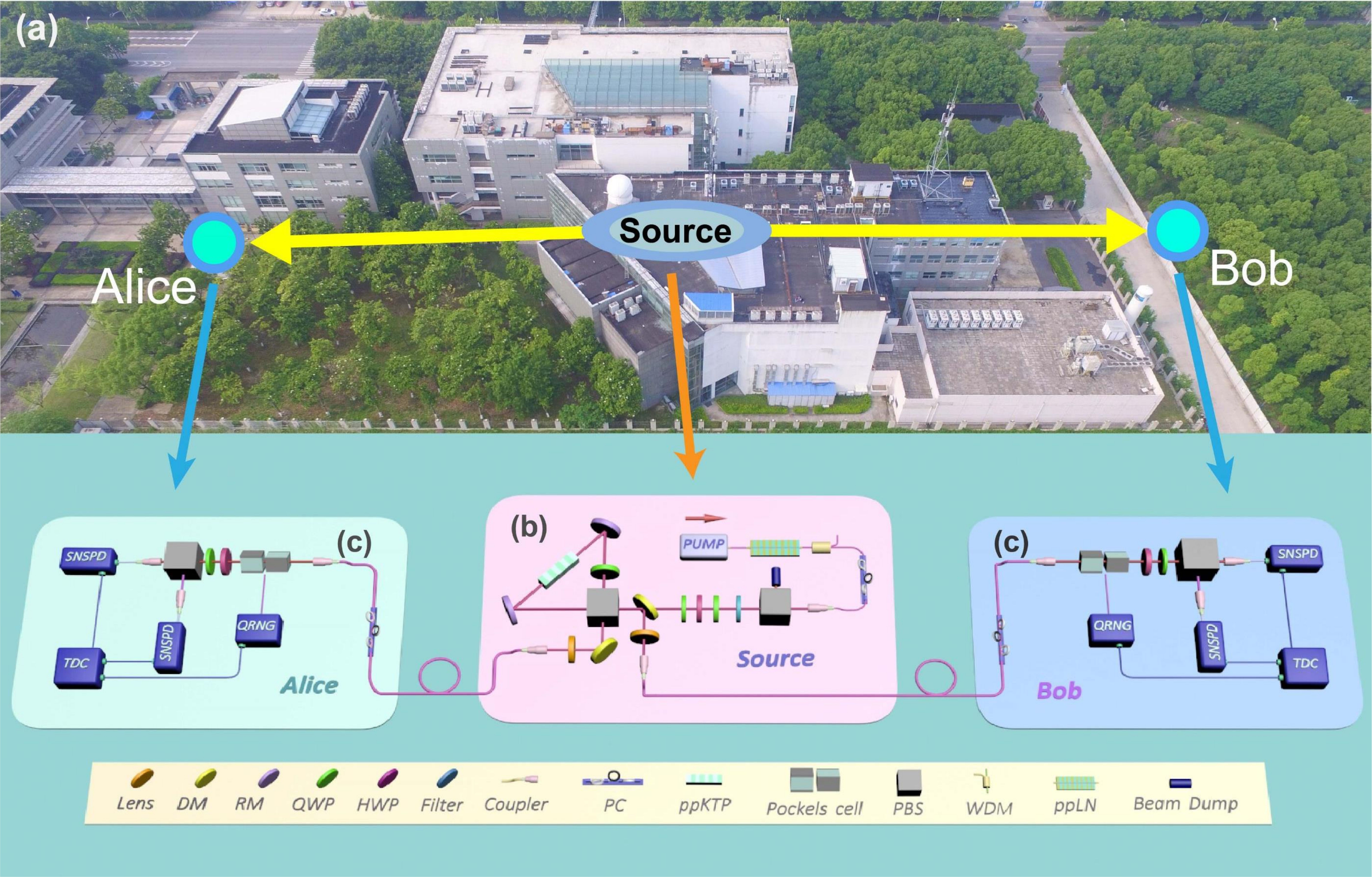}
\caption{Schematics of the experiment for loophole-free Hardy's paradox. (a) Bird's-eye view of the experimental apparatus: Alice and Bob are located on opposite sides of the entanglement source, and the straight-line distance between Alice(Bob) and the source is $93\pm1(90\pm1)$ m. (b) Preparation of entangled photon pairs: Light pulses with a duration of 10 ns and a repetition rate of 200 kHz, generated by a 1560 nm seed laser diode (LD), undergo amplification through an erbium-doped fiber amplifier (EDFA). Subsequently, the pulses are frequency-doubled using an in-line periodically poled lithium niobate (PPLN) crystal. The remaining 1560 nm light is eliminated by a wavelength-division multiplexer (WDM) and spectral filters. Two 780 nm quarter-wave plates (QWPS) and a half-wave plate (HWP) before the Sagnac loop are used to control the polarization of the pump laser thereby changing relative amplitude and phase of the created polarization-entangled photon state.Then the 780nm pump photons are fed into the periodically poled potassium titanyl phosphate (PPKTP) crystal in the Sagnac loop consist of two reflection mirrors (RMs) and a dual-wavelength polarizing beam splitter (PBS) to generate the polarization-entangled photon pairs of $1560$ nm.  After the Sagnac loop, we use dichroic mirrors (DM) to remove the residual 780 nm pump laser. Then the entangled photons are collected into optical fibers by coupler and transferred to Alice and Bob in the opposite sites for polarization projection and measurements. (c) Single-photon polarization measurement: At the measurement side, the photons pass through the fiber, and then undergo polarization state measurements. The setup for performing single-photon polarization measurements comprises a Pockels cell, QWP, HWP, and PBS, and are finally collected into a single-mode optical fiber for detection by superconducting nanowire single-photon detectors (SNSPDs).  There are two SNSPDs at each side to collect the photons transmitted and reflected at the PBS. The measurement settings choice is performed under a Quantum Random Number Generator (QRNG), which is triggered by a 200 kHz signal and generates a random output with a 1:1 ratio of 0 and 1. Here, 0 corresponds to a low voltage and 1 corresponds to a high voltage. After the random signal is generated, it is applied to the Pockels cell, which modifies the polarization measurement by exerting different influences on polarization at different voltage levels. The time-digital convertor (TDC) is applied to keep track of the photon detection and random number generation events.}
\label{setup}
\end{figure*}

\begin{figure}[htbp]
\centering
\includegraphics[width =0.45\textwidth]{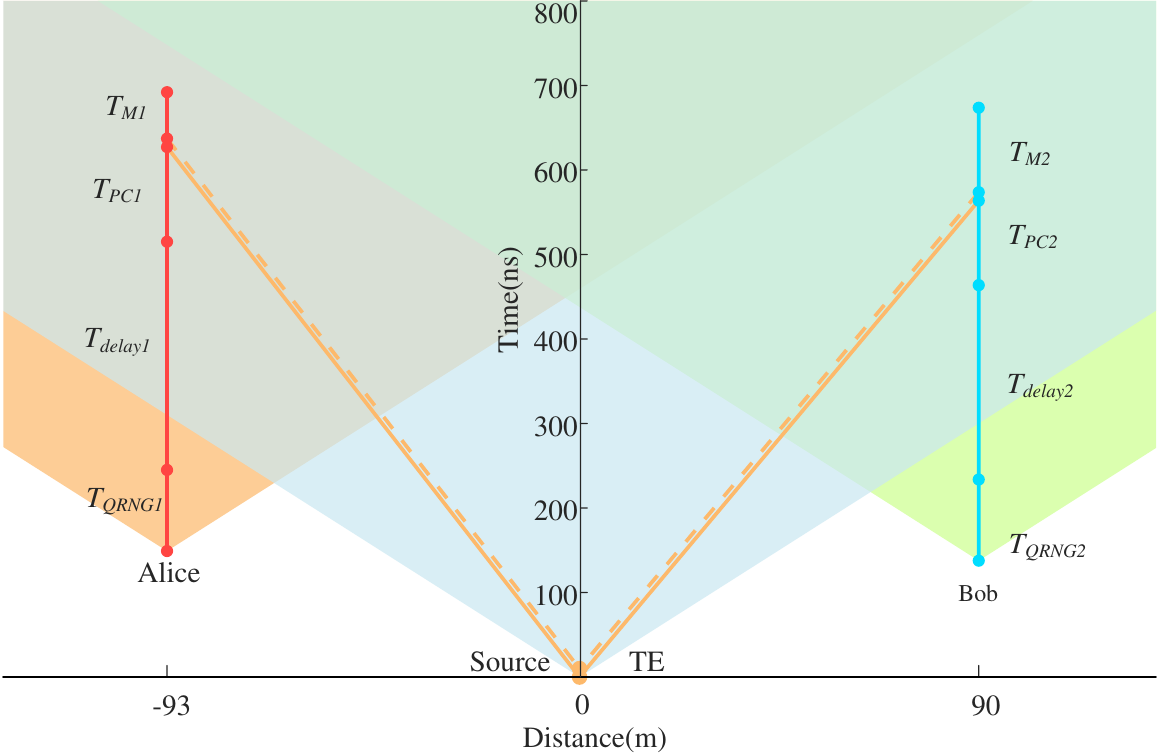}
\caption{Space-time diagram for the experimental events. The red (blue) bar and dots represent the crucial time and node for Alice (Bob)'s measurement. The coordinate origin denotes the beginning
of the emission of photon pairs. The orange line represents the space-time relationship of photons propagation in the optical fiber (The solid orange line represents the propagation of the photons generated at the onset of the generation, while the dashed orange line represents the photons at the end of the generation).  $T_{\text{E}}=10$ ns is the duration of generating entangled photon pairs. $T_{\text{QRNG 1,2}}$ are the durations for QRNG to generate random bits to control the Pockels cells. $T_{\text{delay 1,2}}$ are the times required between the random bits' generation and transferred to Pockels cells. $T_{\text{PC 1,2}}$ are the preparing time for the Pockels cells to be ready for projection measurements after obtaining random bits from QRNGs. $T_{\text{M 1,2}}$ are the duration for SNSPDs to output electric signals. $T_{\text{QRNG 1}}=T_{\text{QRNG 2}}=96$ ns, $T_{\text{delay 1}}=270$ ns, $T_{\text{delay 2}}=230$ ns, $T_{\text{PC 1}}=112$ ns, $T_{\text{PC 2}}=100$ ns, $T_{\text{M 1}}=55$ ns, $T_{\text{M 2}}=100$ ns. The linear distance between Alice (Bob) and the source is $93\pm1$ ($90\pm1$) m, and the corresponding fiber length is $129$ ($116$) m.}
\label{spacetime}
\end{figure}

\mysection{Experiments}\label{experiments}
The present Hardy's paradox test is illustrated in Fig.~\ref{setup}. With the pump laser of 780 nm, the polarization-entangled 1560 nm photon pairs are generated through spontaneous down conversion(SPDC) in the PPKTP crystal within a Sagnac loop. Then the two photons of a pair are transmitted to Alice and Bob's laboratories through fiber links for measurements. To close the \emph{locality loophole} and deal with the \emph{freedom-of-choice loophole}~\cite{Shalm2015Strong,Giustina2015Significant,hensen2015loophole}, we design a space-time configuration for our system, which is shown in Fig.~\ref{spacetime}. Specifically, it's necessary to space-like separate the setting choices on one side (the first dots on the red bar and blue bar denote the beginning of setting choice) from the measurements output
%这里应该没问题，我改的Anton文章里的话，我加了个output
on the other side (the last dots on the two bars denote the ending of the measurement), as well as from the emission of pump photons (the coordinate origin denotes the beginning of the emission of photons and the second dots on the two bars denote the ending of setting choice) which also can be seen as the emission of the hidden variable $\lambda$. The synchronization of the experimental system is achieved by locking the seed laser for the pump light, the QRNG for the setting choice and the time-digital convertor (TDC) for the signal collection to the same clock. We carefully adjust the relative delay between the QRNG and pump light to ensure that the polarization of photons could be modulated when the setting choice signal is applied to the Pockels cell at the same time. Meanwhile, the Alice's measurement station is separated far apart from the Bob's measurement station (93 m for Alice from the source to her laboratory, and 90 m for Bob's case), and the length of optical fibers (129 m for Alice and 116 m for Bob)  are set appropriately  to ensure a time delay of 627 ns and 563 ns respectively from photons' generation to detection.

Furthermore, as for the \emph{freedom-of-choice loophole}, we need to make sure that the random numbers used for measurement setting choices are completely undisturbed and truly random, i.e. the local hidden variables can not manipulate the generation of random numbers. In principle, because that there is an overlap between the backward light cones of two QRNGs and pump laser, the independence and randomness of Alice and Bob's random numbers cannot be proven without making any assumptions~\cite{Shalm2015Strong}. Here, the high speed QRNGs we used could generate random numbers within a time interval of 100 ns after receiving the trigger signal, and the delay time can be flexibly adjusted to accommodate our space-time configuration. 
%\textcolor{blue}{By slightly abusing the assumptions on the randomness and independence in the generation of random numbers and believing the intrinsic randomness of quantum mechanics.
%the choices for measurement settings are made with fast QRNGs in Alice and Bob's laboratories. Meanwhile, the generation events of random numbers are kept outside of the light cone of emission events of entangled photon pairs from the entangled source to avoid the influence of potential hidden variables. 
A detailed space-time analysis in Fig.~\ref{spacetime} shows that \emph{locality loophole} and \emph{freedom-of-choice loophole} can be simultaneously addressed by slightly abusing the assumption on the independence in the generation of random numbers. 

The measurements of Alice and Bob consist of Pockels cells, half-wave plates (HWP),  polarizing beam splitter (PBS) in turn and finally superconducting nanowire single-photon detectors (SNSPD). To distinguish the result 0, 1, and $u$ experimentally, we place two detectors on two output ports of the PBS and label the clicks on transmission and reflection paths as 0 and 1, respectively. Besides, we denote the undetected events (absence of detector clicks) and double-click events (i.e., clicks on both detectors) at one station as $u$. The system heralding efficiencies are measured to be  $82.1\%\pm 0.2\%$ ($82.4\%\pm 0.2\%$) of transmission (reflection) path for Alice, and $82.1\%\pm 0.2\%$ ($82.2\%\pm 0.2\%$) of transmission (reflection) path for Bob, using the SNSPDs with efficiency higher than $96\%$. 
The heralding efficiencies are determined by the ratio of twofold coincidence events to single counts, which corresponds to the total events detected by a single detector directly measured across the entire system without accounting for any losses. These efficiency significantly surpass the record values in previous loophole-free Bell tests with photons (see Table.~\ref{tab:eff}). Furthermore, the efficiencies of two paths at each measurement site are adjusted to be close, as any discrepancy in efficiencies can be interpreted as a change of the measurement bases (because this offset affects the ratio of the probabilities of detecting photons in the transmission and reflection paths). And we employ Hardy's inequality of the form Eq.~\ref{inequ} which has taken all detection events into consideration and inherently closes this loophole. %, which are higher than the minimal detection efficiency. 

\begin{table}[]
    \centering
    \caption{Efficiencies and fidelity in existing photonic experiments of loophole-free Bell tests and related applications. The efficiencies are averaged over Alice’s and Bob’s  detection efficiency.} 
    \label{tab:eff}
    \begin{tabular}{c|c c c |c c}
    \hline
    \hline
      Label&Experiment&Year&Type&Efficiency&Fidelity  \\
      \hline
      (1)&Shalm \textit{et al.}~\cite{Shalm2015Strong}&2015&Bell test&75.15\% \\
      (2)&Giustina \textit{et al.}~\cite{Giustina2015Significant}&2015&Bell test&77.40\%\\
      (3)&Bierhorst \textit{et al.}~\cite{bierhorst2018experimentally}&2018&QRNG&75.50\%\\
      (4)&Liu \textit{et al.}~\cite{liu2018device}&2018&QRNG&78.65\%\\
      (5)&Li \textit{et al.}~\cite{Li2018Test}&2018&Bell test&78.75\%&98.66\%\\
      (6)&Zhang \textit{et al.}~\cite{Zhang2020Experimental}&2020&QRNG&76.00\%\\
      (7)&Shalm \textit{et al.}~\cite{shalm2021device}&2021&QRNG&76.30\%\\
      (8)&Li \textit{et al.}~\cite{Li2021Experimental}&2021&QRNG&81.35\%\\
      (9)&This work&2023&Hardy test&82.22\%&99.10\%\\
      \hline
    \hline
    \end{tabular}

\end{table}

\begin{figure}[htbp]
\centering
\includegraphics[width =0.48\textwidth]{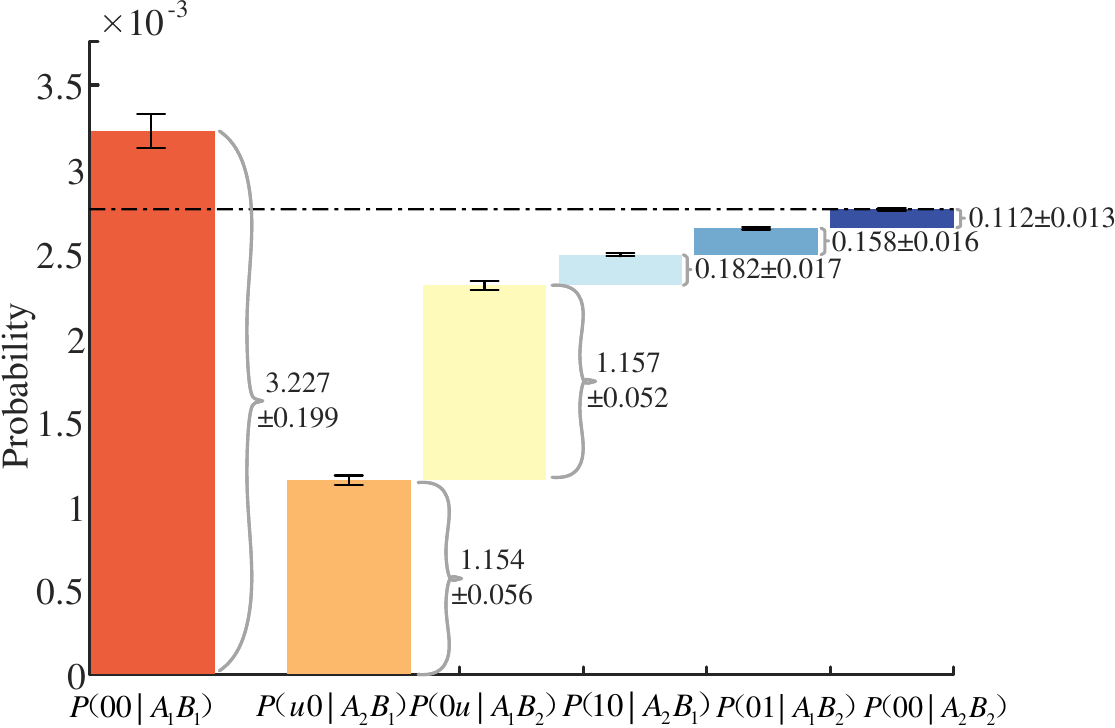}
\caption{Bar chart of the six joint probabilities for the present Hardy's paradox. The height of each bar shows the value of the corresponding joint probability. The last three bars represent three Hardy's conditions, which are small quantities relative to the other probabilities. For the height of the first bar representing $P(00|A_1B_1)$ higher than the sum of the other five bars, the $P_{\text{Hardy}}$ is positive. These results present strong evidence against local realism}\label{resultscomparison}
\end{figure}

To observe the Hardy's nonlocality in this experiment, the quantum state $|\psi(\theta)\rangle=\cos(\theta)\vert HV\rangle+\sin(\theta)\vert VH\rangle$ and measurement settings ${A}_i=\cos(\theta_{A_i})\sigma_z+\sin(\theta_{A_i})\sigma_x$, ${B}_j=\cos(\theta_{B_j})\sigma_z+\sin(\theta_{B_j})\sigma_x$, $i,j\in{1,2}$ are pre-optimized for the overall efficiency $\eta_A(\eta_B)$. Specifically, in the optimization, we set the detection efficiency to be $\eta=82\% $, the corresponding quantum state, Alice and Bob's measurement settings are optimized to be 
$\theta=0.2764$, $\{\theta_{A_1}=-2.8417, \theta_{A_2}=2.1628\}$ and $\{\theta_{B_1}=0.2999, \theta_{B_2}=-0.9788\}$ in radian, respectively.
(See Appendix.~\ref{Optimal_det_effi_and_state} for details). 
In the implementation, we measure the visibility to be 99.5\% and 98.4\% in horizontal/vertical basis and diagonal/antidiagonal basis, respectively. Further, we characterize quantum state by the state tomography measurement, while the fidelity of the non-maximally polarization-entangled state is $99.10\%$ (see See Appendix.~\ref{Experimental Details} for more details). To reduce the dark count, both the window in which Alice and Bob record detection events is set to 15 ns, which is centered on the expected arrival time of Alice and Bob's photons. The average dark count in this window is less than 5 counts per second. 
%这里这个window想说的就是TDC的开门时间，我把TDC删了是不是好一点
After a stable execution of $6$ hours with $4.32 \times 10^{9}$ trials, as shown in Fig.~\ref{resultscomparison}, the observed probabilities in the Hardy's paradox test are $\epsilon_1=P(00|A_2B_2)=(1.120\pm0.136)\times 10^{-4}$,  $\epsilon_2=P(01|A_1B_2)=(1.578\pm0.162)\times 10^{-4}$, $\epsilon_3=P(10|A_2B_1) =(1.818\pm0.171)\times10^{-4}$, $P(0u|A_1B_2)=(1.157\pm0.052)\times10^{-3}$, $P(u0|A_2B_1)=(1.154\pm0.056)\times10^{-3}$, 
and $P(00|A_1B_1)=(3.227\pm0.199)\times10^{-3}$. These probabilities result in a positive Hardy's value $P_{\text{Hardy}}=4.646\times10^{-4}$ based on the Eq.~\ref{inequ}, which is more than $5$ standard deviations according to the observed statistics (here the standard deviation is $\sigma=7.771\times10^{-5}$).
%这里我想说我们得到的epsilon很小，说明是hardy佯谬的验证，而破坏了不等式说明我们是关闭了detection loophole的非定域性验证

To quantify the statistic significant of the Hardy's violation, we conduct a hypothesis test of local realism with the prediction-based ratio (PBR) method of analysis which allows to analyze the experimental results without assuming independent and identical conditions~\cite{Zhang2011Asymptotically}. The null hypothesis is that the experimental results can be accounted for by local hidden variable models. The maximal probability that the observed experimental results comply with the null hypothesis is quantified by a statistical $p$ value. Under the PBR analysis, our demonstration show that the $p$ value is upper bounded by $10^{-16348}$, which is extremely strong evidence against local realism (See Appendix.~\ref{Experimental results} for more details).

\mysection{Discussion and Conclusion}
We have presented a refinement over the theoretical analysis on loophole-free demonstration of Hardy's paradox. By applying high detection efficiency, high fidelity entangled photon source, fast QRNGs, and space-like separating the events corresponding to measurement setting choices, entangled state preparations and photon detections, we have simultaneously closed the \emph{locality loophole}, and addressed the \emph{freedom-of-choice loophole} in the experiment. Combining the strategy of treating double-click events to be inconclusive events and the local hidden variable model under imperfect detection efficiencies, the \emph{detection loophole} is also closed in our experiment at the mean time. 

After a stable execution of $6$ hours, a positive Hardy's value $P_{\text{Hardy}}=4.646\times10^{-4}$ is observed up to more than $5$ standard deviations. Based on a null hypothesis test, the $p$ value that the possibility our results can be explained by local realistic theories doesn't exceed $10^{-16348}$. Therefore, our experiment shows significant evidence that quantum mechanical description on physical quantity cannot be accounted for by local realism. Besides this fundamental interest, these results help to achieve a milestone for quantum information applications based on Hardy's paradox.

\mysection{Acknowledgments}
S.-R. Z., S. Z. and H.-H. D. contributed equally to this work. This work has been supported by the National Natural Science Foundation of China (Grants No.~T2125010, No.~62031024, No.~62375252 and No.~11875167), the National Key R$\&$D Program of China (Grant No.~2019YFA0308700), the Anhui Initiative in Quantum Information Technologies (Grant No.~AHY060200) as well as the Innovation Program for Quantum Science and Technology (Grant No.~2021ZD0301100). J. L. C. was supported by the National Natural Science Foundations of China (Grant Nos.~12275136 and ~12075001) and the 111 project of B23045. S. Z. is supported by the Research Startup Foundation of Hangzhou Dianzi University (No.~KYS275623071) and Zhejiang Provincial Natural Science Foundation of China under Grant No.~LQ24A050005.

\begin{appendix}
\onecolumngrid
\section{Optimal quantum state and measurement settings under imperfect detection efficiency}
\label{Optimal_det_effi_and_state}

At the beginning of the experiment, we optimize the quantum state and measurement settings by considering imperfect system detection efficiency $\eta\in [0,1]$. Following the formalism in the original Hardy's paradox~\cite{hardy1993nonlocality}, the two-qubit entangled state is modeled as 
\begin{equation}\label{app_state}
  |\Psi(\theta)\rangle=\cos(\theta)|01\rangle+\sin(\theta)|10\rangle,
\end{equation}
where $\theta \in [0,\frac{\pi}{2}]$ is an adjustable parameter. When $\theta=\frac{\pi}{4}$, the quantum state becomes a maximal entangled state. For Alice and Bob's binary observables $\{{A}_1,{A}_2\}$ and $\{{B}_1,{B}_2\}$, corresponding projectors are
\begin{equation}
  \begin{split}
    \Pi_{A_i}^a&=|A^a_i\rangle\langle A^a_i|,\\
    \Pi_{B_j}^b&=|B^b_j\rangle\langle B^b_j|.
  \end{split}
\end{equation}
with $i, j\in\{1,2\}$ and $a,b \in\{0,1\}$. Firstly, we take the basis vectors for Alice and Bob's measurements ${A}_2$ and ${B}_2$ to be
\begin{equation}\begin{split}\label{A2B2_1}
  |A^0_2\rangle&=\beta_2|0\rangle+\alpha_2|1\rangle,\\
  |A^1_2\rangle&=\alpha_2|0\rangle-\beta_2|1\rangle,\\
  &\text{and}\\
  |B^0_2\rangle&=-\alpha_2|0\rangle+\beta_2|1\rangle,\\
  |B^1_2\rangle&=\beta_2|0\rangle+\alpha_2|1\rangle.
\end{split}
\end{equation}
Without loss of generality, we have taken amplitudes $\alpha_2$ and $\beta_2$ to be real. Then, the state in the Eq.~\ref{app_state} can be rewritten as
\begin{equation}\label{state1}
  \begin{split}
    |\Psi(\theta)\rangle=&[\cos(\theta)\beta_2^2-\sin(\theta)\alpha_2^2]|A^0_2\rangle|B^0_2\rangle+[\cos(\theta)\alpha_2\beta_2+\sin(\theta)\alpha_2\beta_2]|A^0_2\rangle|B^1_2\rangle\\
                        &+[\cos(\theta)\alpha_2\beta_2+\sin(\theta)\alpha_2\beta_2]|A^1_2\rangle|B^0_2\rangle+[\cos(\theta)\alpha_2^2-\sin(\theta)\beta_2^2]|A^0_2\rangle|B^0_2\rangle.
  \end{split}
\end{equation}
By taking the first term of Eq.~\ref{state1} to be $$\cos(\theta)\beta_2^2-\sin(\theta)\alpha_2^2=0,$$ we have
\begin{equation}
  k^2 \equiv  \frac{\alpha_2^2}{\cos(\theta)}=\frac{\beta_2^2}{\sin(\theta)}.
\end{equation}
Considering $\alpha_2^2+\beta_2^2=1$, $k$ is taken to be $k=\frac{1}{\sqrt{\cos(\theta)+\sin(\theta)}}$ and $\alpha_2=k\sqrt{\cos(\theta)}, \beta_2=k\sqrt{\sin(\theta)}.$ Thus, the first Hardy condition is satisfied, 
\begin{equation}
  P(00|A_2B_2)=Tr(\Pi^0_{A_2}\otimes\Pi^0_{B_2}|\Psi\rangle\langle\Psi|)=0.
\end{equation}
Here, the $k$, $\alpha_2$, and $\beta_2$ have been taken as positive parameters without loss of generality.
Then, the state in the Eq.~\ref{state1} becomes
\begin{equation}\label{state2}
  \begin{split}
    |\Psi(\theta)\rangle &=\sqrt{\sin(\theta)\cos(\theta)}|A^0_2\rangle|B^1_2\rangle+\sqrt{\sin(\theta)\cos(\theta)}|A^0_2\rangle|B^1_2\rangle+[\cos(\theta)-\sin(\theta)]|A^1_2\rangle|B^1_2\rangle\\
                          =&[\frac{\sqrt{\sin(\theta)\cos(\theta)}}{\sqrt{\cos(\theta)-\sin(\theta)}}|A^0_2\rangle+\sqrt{\cos(\theta)-\sin(\theta)}|A^1_2\rangle]\cdot[\frac{\sqrt{\sin(\theta)\cos(\theta)}}{\sqrt{\cos(\theta)-\sin(\theta)}}|B^0_2\rangle+\sqrt{\cos(\theta)-\sin(\theta)}|B^1_2\rangle]\\
                          &-\frac{\sin(\theta)\cos(\theta)}{\cos(\theta)-\sin(\theta)}|A^0_2\rangle|B^0_2\rangle.
  \end{split}
\end{equation}
Now, we choose the observable basis vectors for Alice and Bob's measurements ${A}_1$ and ${B}_1$ to be
\begin{equation}\label{A1B1_1}
  \begin{split}
    |A^0_1\rangle&=-\beta_1|A^0_2\rangle+\alpha_1|A^1_2\rangle,\\
    |A^1_1\rangle&=\alpha_1|A^0_2\rangle+\beta_1|A^1_2\rangle,\\
    &\text{and}\\
    |B^0_1\rangle&=-\beta_1|B^0_2\rangle+\alpha_1|B^1_2\rangle,\\
    |B^1_1\rangle&=\alpha_1|B^0_2\rangle+\beta_1|B^1_2\rangle.
  \end{split}
\end{equation}
with $\alpha_1=\frac{\sqrt{\cos(\theta)\sin(\theta)}}{\sqrt{1-\cos(\theta)\sin(\theta)}}$,$\beta_1=\frac{\cos(\theta)-\sin(\theta)}{\sqrt{1-\cos(\theta)\sin(\theta)}}$ and $N=\frac{1-\cos(\theta)\sin(\theta)}{\cos(\theta)-\sin(\theta)}$. Then, the state in the Eq.\ref{state2} becomes
\begin{equation}
  |\Psi\rangle=N(|A^1_1\rangle|B^1_1\rangle-\alpha_1^2|A^0_2\rangle|B^0_2\rangle).
\end{equation}

Then, in measurement settings $A_2B_2$, $A_1B_2$, $A_2B_1$ and $A_1B_1$, respectively, the quantum state can be rewritten as follows
\begin{subequations}
  \begin{equation}\label{A2B2}
    |\Psi\rangle=N(\alpha_1\beta_1|A^0_2\rangle|B^1_2\rangle+\alpha_1\beta_1|A^1_2\rangle|B^0_2\rangle+\beta_1^2|A^1_2\rangle|B^1_2\rangle),
  \end{equation}
  \begin{equation}\label{A1B2}
    |\Psi\rangle=N[|A^1_1\rangle(\alpha_1|B^0_2\rangle+\beta_1|B^1_2\rangle)-\alpha_1^2(-\beta_1|A^0_1\rangle+\alpha_1|A^1_1\rangle)|B^0_2\rangle],
  \end{equation}

  \begin{equation}\label{A2B1}
    |\Psi\rangle=N[(\alpha_1|A^0_2\rangle+\beta_1|A^1_2\rangle)|B^1_1\rangle-\alpha_1^2|A^0_2\rangle(-\beta_1|B^0_1\rangle+\alpha_1|B^1_1\rangle)],
  \end{equation}

  \begin{equation}\label{A1B1}
    |\Psi\rangle=N[|A_1^1\rangle|B^1_1\rangle-\alpha_1^2(-\beta_1|A^0_1\rangle+\alpha_1|A^1_1\rangle)(-\beta_1|B^0_1\rangle+\alpha_1|B^1_1\rangle)]
  \end{equation}
\end{subequations}
From the Eq.\ref{A2B2}, Eq.\ref{A1B2} and Eq.\ref{A2B1}, we have $P(00|A_2B_2)=0$, $P(01|A_1B_2)=0$, and $P(10|A_2B_1)=0$, respectively. While, from the Eq.\ref{A1B1}, we have 
\begin{equation}\label{ori_Hardy_value_1}
    P(00|A_1B_1)=|\langle A^0_1B^0_1|\Psi\rangle|^2=|N\alpha_1^2\beta_1^2|^2.
\end{equation} 
The Hardy's value $ P(00|A_1B_1)$ can achieve a non-zero value. By considering that ${A}_i=\Pi_{A_i}^0-\Pi_{A_i}^1$ and ${B}_j=\Pi_{B_j}^0-\Pi_{B_j}^1$, the Eq.~\ref{A2B2_1} and Eq.~\ref{A1B1_1} present the solution for the original Hardy's paradox with Hardy's value in Eq.~\ref{ori_Hardy_value_1} under the quantum state Eq.~\ref{app_state}. This certifies the nonlocality through the original Hardy's paradox.

To accommodate the situation with imperfect detection, we assume that detectors in Alice and Bobs' laboratories are symmetric with efficiency $\eta$. When considering imperfect system detection efficiency $\eta$, we would like to claim that the zero Hardy conditions $P(00|A_2B_2)=0$, $P(01|A_1B_2)=0$, and $P(10|A_2B_1)=0$ still hold. The difference is that there will occur undetected events $u$ in both Alice and Bob's laboratories. Inspired by the Ref.~\cite{HWANG1996The}, this situation can be analyzed with the help of the LHV model presented by Eberbard~\cite{Eberhard1993Background}. Based on the above analysis, the observable basis vectors are
\begin{equation}
  \begin{split}
    |A^0_2\rangle&=\frac{1}{\sqrt{\cos(\theta)+\sin(\theta)}}(\sqrt{\sin(\theta)}|0\rangle+\sqrt{\cos(\theta)}|1\rangle),\\
    |A^1_2\rangle&=\frac{1}{\sqrt{\cos(\theta)+\sin(\theta)}}(\sqrt{\cos(\theta)}|0\rangle-\sqrt{\sin(\theta)}|1\rangle),
\end{split}
\end{equation}
and 
\begin{equation}
    \begin{split}
    |B^0_2\rangle&=\frac{1}{\sqrt{\cos(\theta)+\sin(\theta)}}(-\sqrt{\cos(\theta)}|0\rangle+\sqrt{\sin(\theta)}|1\rangle),\\
    |B^1_2\rangle&=\frac{1}{\sqrt{\cos(\theta)+\sin(\theta)}}(\sqrt{\sin(\theta)}|0\rangle+\sqrt{\cos(\theta)}|1\rangle),
\end{split}
\end{equation}
and
\begin{equation}
    \begin{split}
    |A^0_1\rangle&=\frac{1}{\sqrt{\cos(\theta)^3-\sin(\theta)^3}}(\sin(\theta)^{3/2}|0\rangle-\cos(\theta)^{3/2}|1\rangle),\\
    |A^1_1\rangle&=\frac{1}{\sqrt{\cos(\theta)^3-\sin(\theta)^3}}(\cos(\theta)^{3/2}|0\rangle+\sin(\theta)^{3/2}|1\rangle),
\end{split}
\end{equation}
and
\begin{equation}
\begin{split}
    |B^0_1\rangle&=\frac{1}{\sqrt{\cos(\theta)^3-\sin(\theta)^3}}(\cos(\theta)^{3/2}|0\rangle+\sin(\theta)^{3/2}|1\rangle),\\
    |B^1_1\rangle&=\frac{1}{\sqrt{\cos(\theta)^3-\sin(\theta)^3}}(-\sin(\theta)^{3/2}|0\rangle+\cos(\theta)^{3/2}|1\rangle).
  \end{split}
\end{equation}
In this case, the angles for Alice and Bob's measurements should be
\begin{equation}\label{angles}
  \begin{split}
    \tan(\theta_{A_2}/2)&=(\frac{\cos(\theta)}{\sin(\theta)})^{1/2},\tan(\theta_{B_2}/2)=-(\frac{\sin(\theta)}{\cos(\theta)})^{1/2},\\
    \tan(\theta_{A_1}/2)&=-(\frac{\cos(\theta)}{\sin(\theta)})^{3/2},\tan(\theta_{B_1}/2)=(\frac{\sin(\theta)}{\cos(\theta)})^{3/2},
  \end{split}
\end{equation}
where e.g. ${A}_i=\cos(\theta_{A_i})\sigma_z+\sin(\theta_{A_i})\sigma_x=\Pi^0_{A_i}-\Pi^1_{A_i}$ and ${B}_j=\cos(\theta_{B_j})\sigma_z+\sin(\theta_{B_j})\sigma_x=\Pi^0_{B_j}-\Pi^1_{B_j}$. When the measurements for Alice and Bob are taken as above, the Hardy's conditions can be directly satisfied
\begin{equation}\label{condition}\begin{split}
  P(00|A_2B_2)=0, P(01|A_1B_2)=0, P(10|A_2B_1)=0.
\end{split}\end{equation}
Further by considering the effect of imperfect detection efficiency, one has,
\begin{equation}\label{undetect_event}
\begin{split}
    |A^a_i\rangle=\sqrt{\eta}|A^{'a}_i\rangle+\sqrt{1-\eta}|A^{'a,u}_i\rangle,\\
    |B^b_j\rangle=\sqrt{\eta}|B^{'b}_j\rangle+\sqrt{1-\eta}|B^{'b,u}_j\rangle,
\end{split}
\end{equation}
here, $|A^{'a,u}_i\rangle$ ($|B^{'b,u}_j\rangle$) are vectors corresponding with un-detected events and orthogonal to  $|A^{'a}_i\rangle$ ($|B^{'b}_j\rangle$). 
Thus, there will be inconclusive events that should be taken into consideration in Hardy's test. According to the Ref.~\cite{HWANG1996The}, the Eberhard inequality can be used to analyze the effect of non-perfect detection for experimental Hardy's test. Specifically, when Hardy's conditions are satisfied, the LHV model should satisfy the following constraints
\begin{equation}
P(00|A_1B_1)-P(0u|A_1B_2)-P(u0|A_2B_1)\leq 0.
\end{equation}
From Eq.~\ref{A1B1}, Eq.~\ref{A1B2}, Eq.~\ref{A2B1} and Eq.~\ref{undetect_event}, we have
\begin{equation}\begin{split}
P(00|A_1B_1)%&=\eta^2|NA^2B^2|^2\\
             &=\eta^2 [\frac{\cos(\theta)\sin(\theta)(\cos(\theta)-\sin(\theta))}{1-\cos(\theta)\sin(\theta)}]^2,\\
P(0u|A_1B_2)%&=\eta(1-\eta)|NA^2B|^2
             &=\eta(1-\eta)\frac{[\cos(\theta)\sin(\theta)]^2}{1-\cos(\theta)\sin(\theta)},\\
P(u0|A_2B_1)%&=\eta(1-\eta)|NA^2B|^2
             &=\eta(1-\eta)\frac{[\cos(\theta)\sin(\theta)]^2}{1-\cos(\theta)\sin(\theta)}.
\end{split}\end{equation}
%Here, uu denotes inconclusive events, ++ and −- denote the eigenvectors for Alice and Bob's observable with +1+1 and −1-1 eigenvalues, respectively.
Defining the following function of $\theta$ and $\eta$,
\begin{equation}\begin{split}
P_{\text{Hardy}}(\theta,\eta)&=P(00|A_1B_1)-P(0u|A_1B_2)-P(u0|A_2B_1).
\end{split}\end{equation}
When the maximal violation of Hardy's paradox is achieved and considering $\theta\in[0,\frac{\pi}{2}]$, there should be
\begin{equation}\begin{split}
\frac{\partial P_{\text{Hardy}}(\theta,\eta)}{\partial \theta}=0  \Rightarrow \theta = \frac{1}{2}\arcsin(3-\sqrt{\frac{6\eta-1}{2\eta-1}}).
\end{split}\end{equation}
Recalling the Eq.~\ref{angles}, this gives the optimal quantum states and observables at given different detection efficiency $\eta$.  The maximal violation at detection efficiency $\eta$ is termed as
\begin{equation}
P_{\text{Hardy}}^{\text{max}}(\eta)=\frac{1}{2}[1-\sqrt{1+4\eta(3\eta-2)}]+3\eta[1-3\eta+\sqrt{1+4\eta(3\eta-2)}].
\end{equation}
For the quantum state in the Eq.~\ref{app_state}, there will be positive Hardy's values $P^{max}_{\text{Hardy}}>0$ only when $\eta>\frac{2}{3}$. In Table~\ref{optim_parameter_performance}, we present the optimal quantum states and observables together with maximal Hardy's values at detection efficiencies around $\eta=82\%$.

\begin{table*}[htbp]
  \centering
  \resizebox{\textwidth}{14mm}{
  \begin{tabular}{|c|c|c|c|c|c|c|c|c|c|c|c|c|}
     \hline
    $\eta$ & $\theta_{A_1}$ & $\theta_{A_2}$ & $\theta_{B_1}$ & $\theta_{B_2}$ & $\theta$ & $P(00|A_2B_2)$ & $P(01|A_1B_2)$ & $P(10|A_2B_1)$ & $P(00|A_1B_1)$ & $P(0u|A_1B_2)$& $P(u0|A_2B_1)$ & $P_{\text{Hardy}}^{\text{max}}(\eta)$ \\
  \hline
  $0.788$ & $-2.90572$ & $2.22849$ & $0.235875$ & $-0.913103$ & $0.236716$ & $0$ & $0$ & $0$ & $0.029457$ & $0.011246$ & $0.011246$ & $0.00696561$ \\
  \hline
  $0.8$ & $-2.88135$ & $2.20203$ & $0.260239$ & $-0.939565$ & $0.252261$ & $0$ & $0$ & $0$ & $0.033588$ & $0.012326$ & $0.012326$ & $0.00893709$ \\
  % \hline
  % $0.805$ & $-2.87116$ & $2.19152$ & $0.270433$ & $-0.950068$ & $0.258597$ & $0$ & $0$ & $0$ & $0.035314$ & $0.012734$ & $0.012734$ & $0.00984483$ \\
  \hline
  $0.81$ & $-2.86136$ & $2.18172$ & $0.280231$ & $-0.959875$ & $0.2646$ & $0$ & $0$ & $0$ & $0.037038$ & $0.013117$ & $0.013117$ & $0.0108044$ \\
  \hline
  % $0.815$ & $-2.85147$ & $2.17208$ & $0.290122$ & $-0.969508$ & $0.270577$ & $0$ & $0$ & $0$ & $0.038761$ & $0.013472$ & $0.013472$ & $0.0118164$ \\
  % \hline
  $0.82$ & $-2.84165$ & $2.16277$ & $0.29994$ & $-0.97882$ & $0.276432$ & $0$ & $0$ & $0$ & $0.040478$ & $0.013798$ & $0.013798$ & $0.0128816$ \\
  \hline
  $0.83$ & $-2.82223$ & $2.14503$ & $0.319365$ & $-0.996562$ & $0.287797$ & $0$ & $0$ & $0$ & $0.043894$ & $0.01436$ & $0.01436$ & $0.0151737$ \\
  \hline
  $0.84$ & $-2.80308$ & $2.12835$ & $0.338509$ & $-1.01324$ & $0.298732$ & $0$ & $0$ & $0$ & $0.047274$ & $0.014794$ & $0.014794$ & $0.0176853$ \\
  \hline
  $0.85$ & $-2.78422$ & $2.11262$ & $0.357376$ & $-1.02897$ & $0.30927$ & $0$ & $0$ & $0$ & $0.050606$ & $0.015093$ & $0.015093$ & $0.0204202$ \\
  \hline
   \end{tabular}
   }
  \caption{Optimal quantum states and measurement setting versus detection efficiency $\eta$. In the optimization, the detection efficiency for detectors in Alice and Bob's laboratories is assumed to be symmetric. The $\theta$, $\theta_{A_i}$ and $\theta_{B_j}$  with $i,j\in{1,2}$ are parameters for the quantum state, Alice's measurement settings and Bob's measurement settings, respectively.}\label{optim_parameter_performance}
\end{table*}

In the experiment, apart from detection efficiency, there exist other system parameters that influence the maximal violation of Hardy’s paradox, among which the fidelity of the state and the average photon number per pulse are of utmost importance. We conducted simulations for these variables as well to guide our parameter selection in the experiment. The entangled photon pairs are prepared using an SPDC source, the pulses of which may include the vacuum and multiple pairs of photons besides single pair of photons. We assume that the probabilities for the component of $n$ pairs of photons in the output of an SPDC source follow the Poisson distribution,
\begin{equation}\label{poisson distribution}
  P(n) = \frac{\mu^n}{n!}e^{-\mu},
\end{equation}
where $\mu$ is the mean photon number in the output of the SPDC source. For each pair of photons, the experimentally prepared quantum state may be imperfect that could deviate from the one defined in Eq.~\ref{app_state}. We use the Werner state to simulate the prepared state. Which can be written as:
\begin{equation}\label{wener_state}
  \rho = V|\Psi(\theta)\rangle\langle\Psi(\theta)|+\frac{1-V}{4}I
\end{equation}
where V is the visibility of the state, and the corresponding fidelity is $F = (3V+1)/4$.

\begin{figure}[htbp]
\centering
\includegraphics[width =0.5\textwidth]{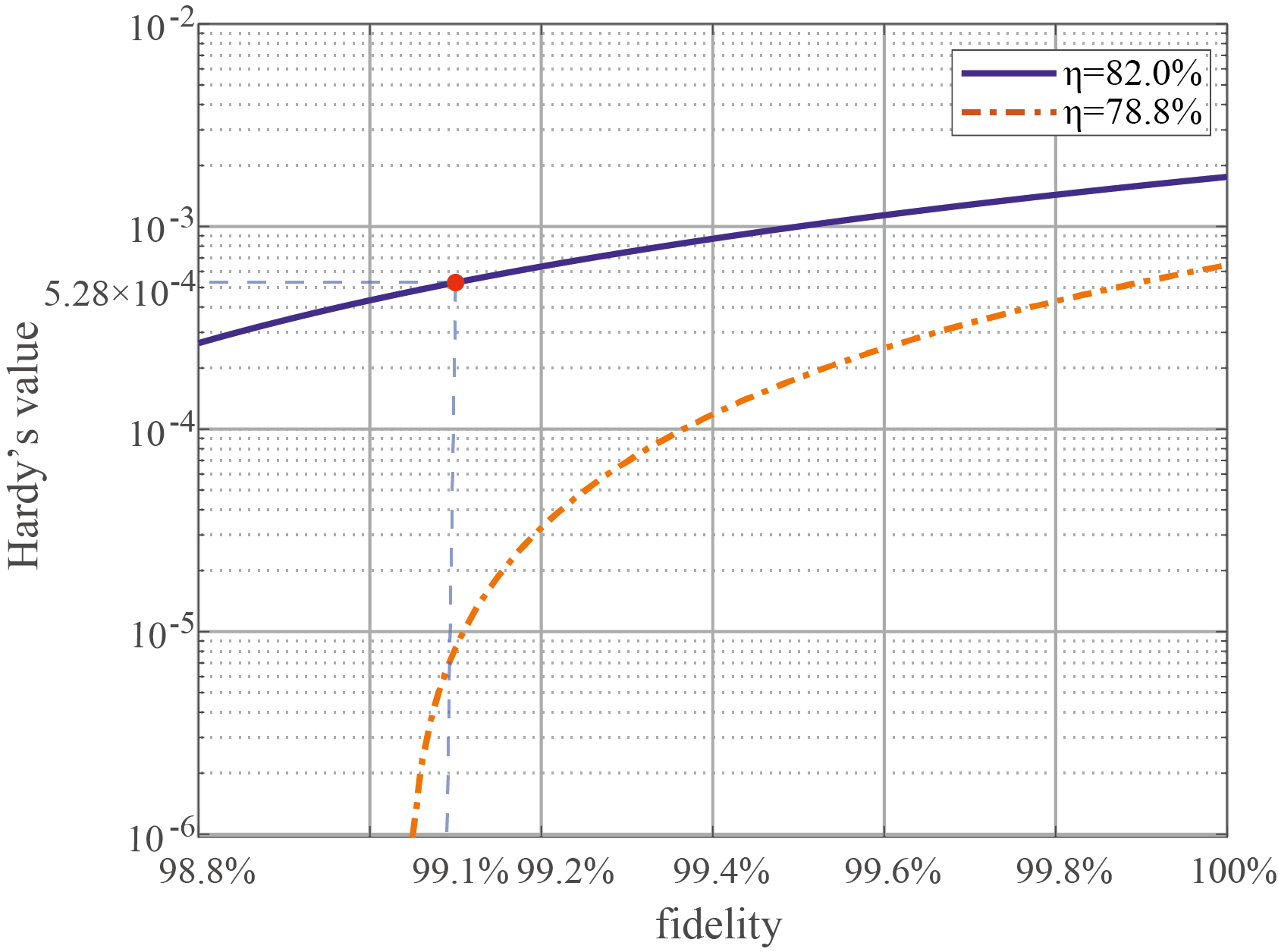}
\caption{Hardy's values versus fidelity for $\eta = 78.8\%$ and $\eta = 82\%$. The red point represent the Hardy's value of $5.28\times10^{-4}$ for our system. }\label{fidelity}
\end{figure}

As shown in Fig.~\ref{fidelity}, we present the relations between Hardy's value and fidelity under detection efficiencies $\eta=78.8\%$ and $82\%$. Here, $\eta=82\%$ is the detection efficiency for parameter optimization in this work, which is slightly lower than the system detection efficiencies in our experiment. $\eta=78.8\%$ is taken from Ref.~\cite{Li2018Test}, which is one of the latest loophole-free Bell inequality tests. From the results of Fig.~\ref{fidelity}, the theoretical Hardy's value should be less than $10^{-6}$ with the detection efficiency (78.88\%) and fidelity (98.66\%) in Ref.~\cite{Li2018Test}. While, for our system, the detection efficiency is around 82\%, and the fidelity is 99.10\%.  We can realise a Hardy's value of $5.28\times10^{-4}$ theoretically (the red point in Fig.~\ref{fidelity}).

\section{Experimental Details}\label{Experimental Details}

\subsection{Modeling the inconclusive events}
Recalling the schematic setup in the main text, there are two threshold single photon detectors on Alice's (Bob's) side. The one on the transmission path after PBS corresponding to the result 0, and the other on the reflection path corresponding to 1. Because of the imperfect detection efficiency, dark counts, the vacuum state and multiple photon pairs in the output of SPDC source, there will be two kinds of inconclusive events that both two detectors on one side click or neither has a click. In this experiment, in order to close the \emph{detection loophole}, we treat both inconclusive events as $u$. 
%If we want to test the CHSH inequality, we will merge the result 0 and result $u$.

%non-photon-number-resolving 

% If there is no dark count, for the vacuum state, there will be no click on the two detectors on each side, then we will only have the result $(uu|A_1B_1), (uu|A_1B_2), (uu|A_2B_1), (uu|A_2B_2),$ which contributes 0 to the Hardy's value and the Eberhard value.
% For the single photon pair condition, we will have result u because of our imperfect detection efficiency, this is the case that the theory of Hwang, Koh and Han Ref.~\cite{HWANG1996The} deal with. 

% As for the multi photon pairs events, this may result in the multi-click on one detector or two detectors on each side. If one detector is clicked multi-times in a trial on Alice or Bob’s side, we record the result as what we get. Because the used photon detectors cannot distinguish the click in one trial, and the photons set to Alice and Bob in two different pairs have no quantum correlations, so the result we record will have a positive contribution to the Hardy's value if and only if this result is product by a pair of entangle photons. As for the other case, both two detectors click on Alice or Bob’s side, we record it as result u, which will only have a negative contribution to the Hardy's value.

\subsection{Quantum state characterization}\label{App:state}

To obtain the maximal Hardy's values in our experiment, the quantum state is set to be the non-maximally polarization entangled two-photon state
$\cos(0.2764)\vert HV\rangle+\sin(0.2764)\vert VH\rangle$ and we set the angles for single
photon polarization state measurement to be $\theta_{A_1}=-2.8417, \theta_{A_2}=-2.1628$ for Alice, $\theta_{B_1}=0.2999, \theta_{B_2}=-0.9788$ for Bob (i.e. the optimized quantum state and measurement settings for $\eta=82\%$ in Sec.~\ref{Optimal_det_effi_and_state}). To create this quantum state in the experiment, we measure diagonal/anti-diagonal visibility in the bases set $(\pi/4,-0.2764)$, $(\pi/2+0.2764,\pi/4)$ for minimum coincidence, and in the bases set $(\pi/4, \pi/2-0.2764)$, $(0.2764, \pi/4)$ for maximum coincidence, where the angles represent measurement basis $\cos(\theta)\sigma_z+\sin(\theta)\sigma_x$ for Alice and Bob. Meanwhile, we perform the state tomography measurement on the non-maximally entangled state. We use the maximum likelihood estimation of density matrices to calculated the fidelity to avoid the problem of experimental inaccuracies and statistical fluctuations of coincidence counts.

\begin{figure}[htbp]
\centering
\includegraphics[width =1\textwidth]{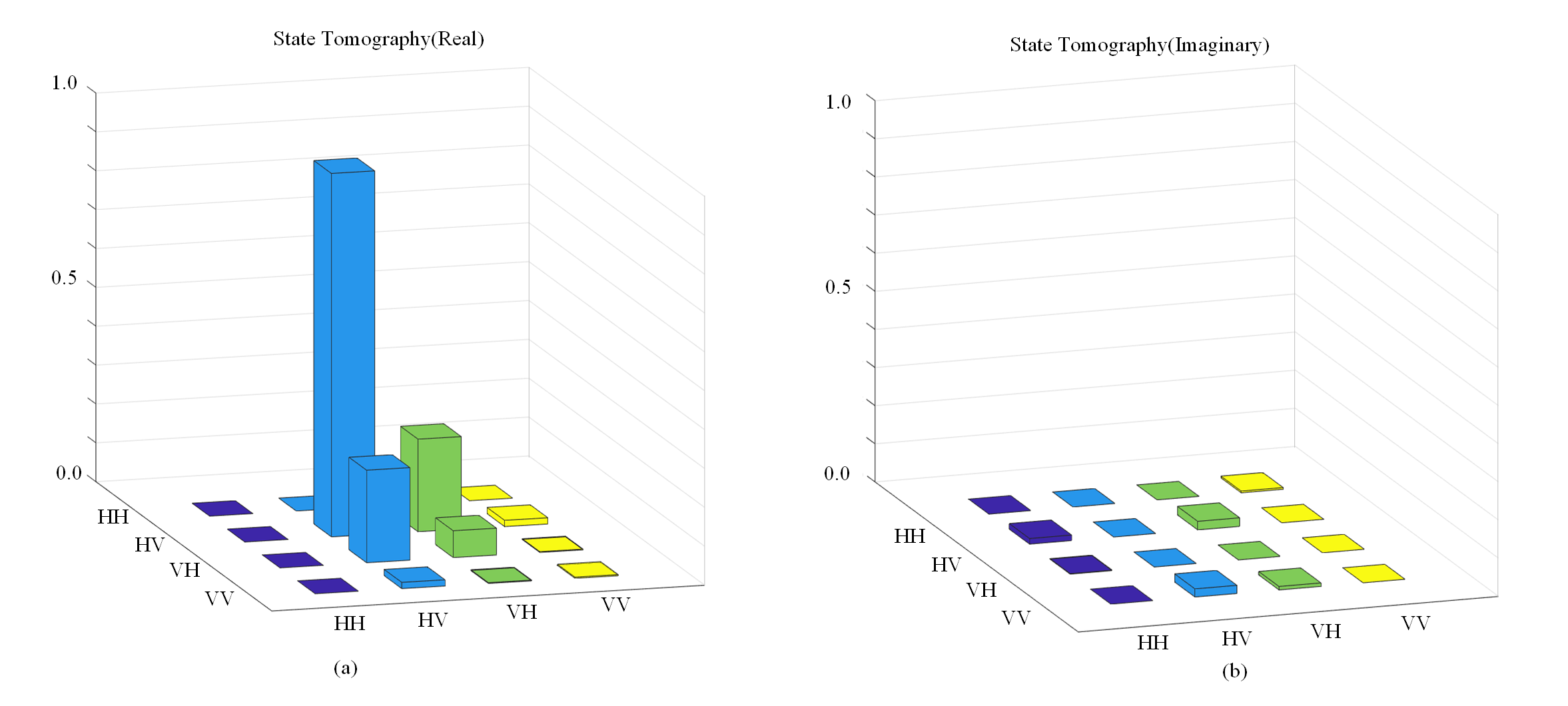}
\caption{(color online) Tomography of the prepared quantum state. The real and imaginary parts are shown in (a) and (b), respectively.}\label{tomo}
\end{figure}

We generate a formula for an explicitly ‘‘physical’’ density
matrix, i.e., a matrix that has the three important properties
of normalization, Hermiticity, and positivity. This matrix will
be a function of 16 real variables with $t=\{t_1 ,t_2 , . . . ,t_{16}\}$ and is denoted as $\rho_p(t)$. For any matrix that can be written in the form $G = T^\dag T$ must be non-negative definite. The explicitly ‘‘physical’’ density matrix $\rho_p$ is given by the formula
\begin{equation}
    \rho_p = T^\dag(t)T(t)/\text{Tr}\{T^\dag(t)T(t)\}
\end{equation}
and it is convenient to choose a tridiagonal form for $T$ :
\begin{equation}
    T(t) = \left[ {\begin{array}{cccc}t_{1} & 0 & 0 & 0 
    \\t_{5}+it_6 & t_{2} & 0 & 0 
    \\t_{11}+it_{12} & t_{7}+it_8 & t_{3} & 0 
    \\t_{15}+it_{16} & t_{13}+it_{14} & t_{9}+it_{10} & t_{4}
    \\\end{array} } \right]
\end{equation}
The measurement data consists of a set of coincidence
counts $n_\mu$ whose expected value is $\overline{n_\mu}=N\langle\phi_\mu|\rho|\phi_\mu\rangle$. Here $\rho$ is the prepared quantum state. In our experiment $\mu = 1,2,...,36$, $|\phi_\mu\rangle\langle\phi_\mu|$ is the operator of the projection measurement of the two-photon state, and for each photon we measured in the bases $|H\rangle,|V\rangle,|+\rangle,|-\rangle,|R\rangle,|L\rangle$, there are 36 projection measurements for $\rho$, $\phi_\mu\in{|HH\rangle,|HV\rangle,......|RL\rangle,|RR\rangle}$. Assuming that the noise on these coincidence measurements has a Gaussian probability distribution. Thus the probability of obtaining a set of 36 counts ${n_1 ,n_2 , . . . n_{36}}$ is 
\begin{equation}
    P(n_1 ,n_2 , . . . n_{36}) = \frac{1}{N_{norm}}\prod\limits_{\mu=1}^{36}\exp[-\frac{(n_{\mu}-\overline{n_\mu})^2}{2\sigma^2_\mu}]
\end{equation}
where $\sigma_\mu$ is the standard deviation for the $n$-th coincidence measurement (given approximately by $\sqrt{\overline{n_\mu}}$) and $N_{norm}$ is the normalization constant. For our physical density matrix $\rho_p$, the number of counts expected for the $n$-th measurement is
\begin{equation}
    \overline{n_\mu}(t_1,t_2,...,t_{16}) = N\langle\phi_\mu|\rho_p(t_1,t_2,...,t_{16})|\phi_\mu\rangle
\end{equation}
Thus the likelihood that the matrix $\rho_p(t_1,t_2,...,t_{16})$ could
produce the measured data $n_1 ,n_2 , . . . ,n_{36}$ is
\begin{equation}
    P(n_1 ,n_2 , . . . n_{36}) = \frac{1}{N_{norm}}\prod\limits_{\mu=1}^{36}\exp[-\frac{(N\langle\phi_\mu|\rho_p(t_1,t_2,...,t_{16})|\phi_\mu\rangle-n_\mu)^2}{2N\langle\phi_\mu|\rho_p(t_1,t_2,...,t_{16})|\phi_\mu\rangle}]
\end{equation}
Rather than to find the maximum value of $P(t_1 ,t_2 , . . . ,t_{16})$, it simplifies things somewhat to find the maximum of its logarithm (which is mathematically equivalent). Thus the optimization problem reduces to finding the minimum of the following function:
\begin{equation}
    L(t_1 ,t_2 , . . . t_{16}) = \sum_{\mu=1}^{36}\frac{(N\langle\phi_\mu|\rho_p(t_1,t_2,...,t_{16})|\phi_\mu\rangle-n_\mu)^2}{2N\langle\phi_\mu|\rho_p(t_1,t_2,...,t_{16})|\phi_\mu\rangle}
\end{equation}
This is the ‘‘likelihood’’ function that we employed in our numerical optimization routine.
The result is shown in
Fig.~\ref{tomo}, the state fidelity is $99.10\%$. The coordinate axis labeled by HH, HV, VH and VV for the density operator$\rho$ can be written as:
\begin{equation}
\rho = \rho_{11}|HH\rangle\langle HH|+\rho_{12}|HH\rangle\langle HV|+...+\rho_{44}|VV\rangle\langle VV|
\end{equation}
where $\rho_{ij}$ if the elements of the density operator matrix. The main imperfections are attributed to the multi-photon components, imperfect optical elements, and imperfect spatial/spectral mode matching.

\subsection{Spacetime configuration of the experiment}\label{App:spacetime_details}

\begin{figure}[htbp]
\centering
\includegraphics[width =0.5\textwidth]{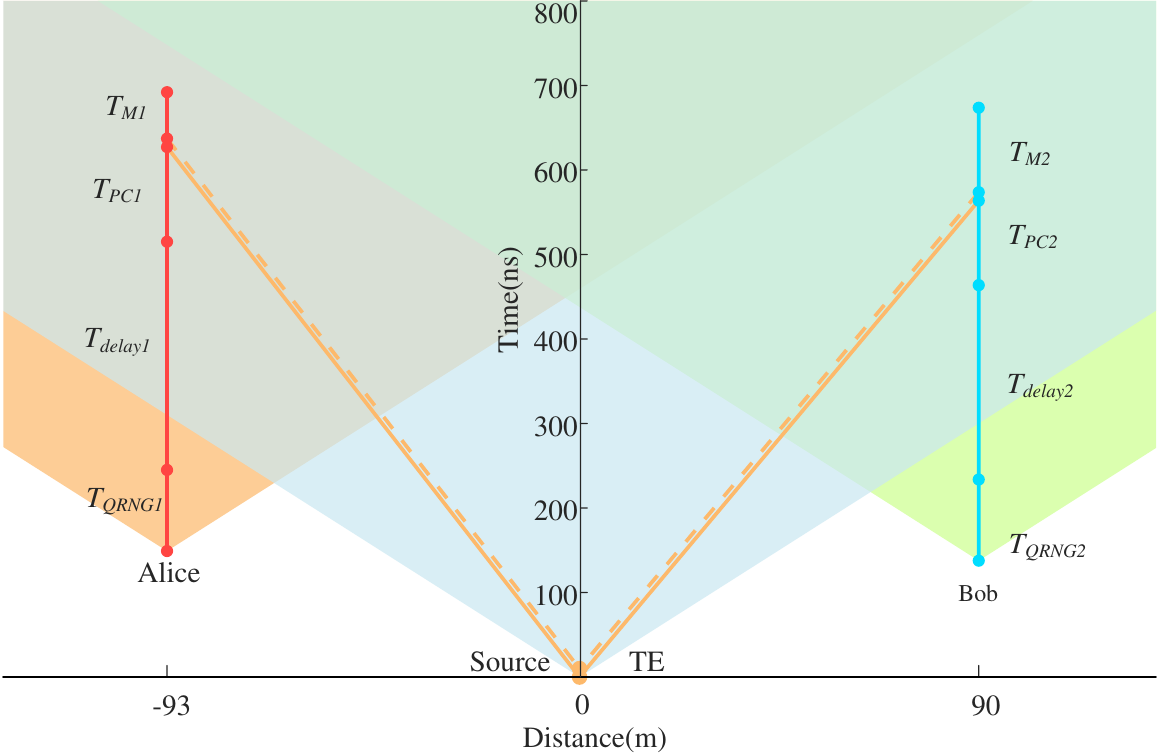}
\caption{Space-time diagram for the experimental events. $T_{E}=10 ~\text{ns}$ is the generation time of entangled photon pairs. $T_{QRNG1,2}$ are duration of times required to generate random bits to switch the Pockels cells. $T_{delay1,2}$ are the delay time between the random bits being generated and received by the Pockels cells. $T_{PC1,2}$ are the waiting times for the Pockels cells to perform state measurements after receiving the random bits. $T_{M1,2}$ are the times taken by the single photon detectors to output electronic signals. $T_{QRNG1} = T_{QRNG2} = 96 ~\text{ns}$, $T_{delay1} = 270 ~\text{ns}$, $T_{delay2} = 230 ~\text{ns}$, $T_{PC1} = 112 ~\text{ns}$,
$T_{PC2} = 100 ~\text{ns}$, $T_{M1} = 55 ~\text{ns}$ and $T_{M2} = 100~ \text{ns}$. Alice’s and Bob’s
measurement stations are placed on opposite sides of the source at distances of 93 m and 90 m, respectively. The effective optical length between Alice’s (Bob’s) station and the source is 132 m (119 m). This arrangement ensures no signalling between relevant events in the experiment. The shaded areas are the future light cones for the source, Alice's and Bob's laboratories.}\label{spacelike}
\end{figure}

To rule out the locality loophole, space-like separation must be satisfied between measurement events at Alice and Bob’s measurement stations, i.e., the setting choice and measurement outcome at one station must be space-like separated from the events at the other station (see Fig.~\ref{spacelike}). Therefore, we then obtain

\begin{equation}
\left\{\begin{array}{l}
(|SA|+|SB|) / c>T_{E}-\left(L_{SA}-L_{SB}\right) / c+T_{QRNG1}+T_{\text {Delay}1}+T_{PC1}+T_{M2} \\
(|SA|+|SB|) / c>T_{E}+\left(L_{SA}-L_{SB}\right) / c+T_{QRNG2}+T_{\text {Delay}2}+T_{PC2}+T_{M1}
\end{array}\right.
\end{equation}
where $|SA| = 93~\text{m}$ $(|SB| = 90~\text{m})$ is the free space distance between entanglement source and Alice’s (Bob’s) measurement station. $T_E = 10~\text{ns}$ is the generation time for photon pairs, which is mainly contributed by the $10~\text{ns}$ pump pulse duration. $L_{SA} = 188~\text{m}$ $(L_{SB} = 169~\text{m})$ is the effective optical path that is mainly contributed by the $129~\text{m} (116~\text{m})$ long fiber between the source and Alice's (Bob’s) measurement station. $T_{QRNG1} = T_{QRNG2} = 96~\text{ns}$ is the time elapse for QRNG to generate a random bit. $ T_{Delay1} = 270~\text{ns}$ $(T_{Delay2} = 230~\text{ns})$ is the delay between the random numbers being generated and delivered to the Pockels cell. $ T_{PC1} = 112~\text{ns}$ $(T_{PC2} = 100~\text{ns})$ is the waiting time for the Pockels cell to be ready to perform state measurements after receiving the random numbers, including the internal delay of the Pockcels Cells $(62~\text{ns}, 50~\text{ns})$ and the time for Pockcels cell to stabilize before performing single photon polarization state projection after switching of $50~\text{ns}$. $T_{M1} = 55~\text{ns}$ $(T_{M2} = 100~\text{ns})$ is the time elapse for SNSPD to output an electronic signal, including the delay due to fiber and cable length (in our experiment, the fiber and cable used in Alice or Bob after the two coupler has the same length. The time difference caused by them is less than $2~\text{ns}$, so we neglect it in analysis). The electronic signals generated by SNSPD are considered clonable and thus represent a definite classical entity. While the photon involved in the entangled state is considered unclonable according to quantum physics. Therefore, we can reasonably assume that the photons collapse upon reaching the SNSPD, thereby concluding the measurement of the photon at the SNSPD. In that sense, the collapse-locality loophole, which concerns where the measurement outcomes eventually arise in space-time, was strongly tighten.

Measurement independence between entangled-pair creation event and setting choice events is satisfied by the following spacelike separation configuration:
\begin{equation}
\left\{\begin{array}{l}
|S A| / c>L_{S A} / c-T_{\text {Delay } 1}-T_{P C 1}, \\
|S B| / c>L_{S B} / c-T_{\text {Delay } 2}-T_{P C 2}.
\end{array}\right.
\end{equation}
We estimate the fiber length by measuring the reflection, the single photon arrives at the SNSPD and generates an electronic response with high efficiency. However, with a small possibility, the photon is reflected by the SNSPD chip, travels to the source, gets polarization rotated in the Sagnac loop, and then travels back to SNSPD, making the second click. We measure the electronic cable length using a ruler. By subtracting the time for the signal passing through the fiber and cable, as well as the delay caused by the discriminator, we estimate the effective fiber length between the Pockels cell and the SNSPD chip. The measured fiber length, cable length, and discriminating time are summarized in Tab.~\ref{tab:length}.

\setlength{\tabcolsep}{8mm}
\begin{table}[ht]
    \centering
    \caption{The fiber distances between Source and Measurement.}
\begin{tabular}{c|ccc}
\hline
      & \textbf{Source-PC} & \textbf{PC-SNSPD} & \textbf{SNSPD-TDC} \\ \hline
\textbf{Alice} & 129 m     & 11 m     & 4 m       \\
\textbf{Bob}   & 116 m     & 20 m     & 12 m      \\ \hline
\end{tabular}
\label{tab:length}
\end{table}

\subsection{Determination of single photon efficiency}

We use the heralding efficiency as the single photon efficiency in our experiment. The heralding efficiency is defined as $\eta_A = C/N_B$ and $\eta_B = C/N_A$ for Alice and Bob, Where C is the coincidence events and $N_A$, $N_B$ is the single events for Alice and Bob measured in the experiment. The heralding efficiency is given by
\begin{equation}
\eta = \eta_{sc} \times \eta_{so} \times \eta_{fiber} \times \eta_m \times \eta_{det}
\end{equation}
where $\eta_{sc}$ is the efficiency to couple entangled photons into single mode optical fiber, $\eta_{so}$ is the efficiency for photons passing through the optical elements in the source, $\eta_{fiber}$ is the transmittance of fiber connecting source to measurement station, $\eta_m$ is the efficiency for light passing through the measurement station, and $\eta_{det}$ is the single photon detector efficiency. We list the heralding efficiencies from the entangled photon source to detection stations, the transmission efficiencies of each intermediate process, and the detection efficiencies of the detectors in Tab.~\ref{tab:efficiency}. We measure the transmission efficiencies of each process and the detection efficiencies with classical light beams and NIST-traceable power metres,

\begin{table}[]
\centering
\caption{The efficiencies of optical elements.}\label{opt}
\begin{tabular}{l|l}
\hline
Optical element                         & Efficiency       \\ \hline
Aspherical lens                         & 99.27\% ± 0.03\% \\
Spherical lens                          & 99.6\% ± 1.0\%   \\
Half wave plate (780~nm/1560~nm)          & 99.93\% ± 0.02\% \\
Half wave plate (1560~nm)                & 99.92\% ± 0.04\% \\
Quarter wave plate (1560~nm)             & 99.99\% ± 0.08\% \\
Polarizing beam splitter (780~nm/1560~nm) & 99.6\% ± 0.1\%   \\
Polarizing beam splitter (1560~nm)       & 99.6\% ± 0.2\%   \\
Dichoric mirror                         & 99.46\% ± 0.03\% \\
PPKTP                                   & 99.6\% ± 0.2\%   \\
Pockels cell                            & 98.7\% ± 0.5\%   \\ \hline
\end{tabular}
\end{table}

\begin{table}
    \centering
    \caption{Photon transmission and detection efficiencies in the experiment.}
    \begin{tabular}{c| c| c c c c c}
    \hline
    Parties & $\eta$ & $\eta_{sc}$ & $\eta_{so}$ & $\eta_{fiber}$ & $\eta_{m}$ & $\eta_{det}$\\
    \hline
    Alice1 & 82.1\% & \multirow{2}{*}{94.3\%} & \multirow{4}{*}{95.9\%} & \multirow{4}{*}{99\%} & {95.5\%} & 96.0\% \\
    Alice2 & 82.4\% & ~ & ~ & ~ & {95.6\%}& 96.3\% \\
    Bob1 & 82.1\% & \multirow{2}{*}{94.0\%} & ~ & ~ &{95.2\%} & 96.7\% \\
    Bob2 & 82.2\% & ~ & ~ & ~ & {95.5\%} & 96.5\% \\
    \hline
    \end{tabular}
    \label{tab:efficiency}
\end{table}

\section{Experimental results}\label{Experimental results}
%\subsection{Statistical analysis of output}
\subsection{Test of local realism}\label{lr}
To convincingly quantify the results of the experiment, we perform a hypothesis test of local realism. The null hypothesis is that the experimental results can be explained by local realism. We use the $p$ value obtained from test statistics to denote the maximum probability that the observed experimental results can be obtained from local realism. Hence, we have to show that our results correspond to a very small $p$ value so that they can indicate a strong rejection of local hidden-variable models. We apply the prediction-based radio (PBR) method to design the test statistics and compute the upper bound of the $p$ value. 

Denote Alice’s and Bob’s random settings distribution at each trial as $p_{A}(x)$ and $p_{B}(y)$ where $x,y\in\{0,1\}$. We assume the joint probability distribution of them $p_{xy}$ is fixed and known before running the test. The measurement outcomes of Alice and Bob at each trial are denoted by A and B with possible value $a,b\in\{0,1,u\}$. In practice, we can only obtain the frequency of different outcomes $\textbf{f}\equiv\{n(abxy)/N,a,b=0,1,u;x,y=0,1\}$, where n is number of trials the result $(a,b,x,y)$ appears and N is the number of the total experimental trials.
So, at the beginning of the test, we have to use the maximum likelihood method to find out the no-signaling distribution $\textbf{P}_{\text{NS}}^{*}\equiv\{p_{xy}p_{\text{NS}}^{*}(ab|xy),a,b=0,1,u;x,y=0,1\}$ that has the minimum distance from the observed frequency distribution $\textbf{f}$. When the trials is large enough, $\textbf{P}_{\text{NS}}^{*}$ can be close to the true probability distribution.

Particularly,  we use the Kullback-Leribler (KL) Ref.~\cite{kullback1951information} divergence to measure this distance because it’s the optimal 
asymptotic efficiency for rejecting the null hypothesis.
\begin{equation}
    D_{\text{KL}}(\textbf{f}||\textbf{P}_{\text{NS}})=\sum_{a,b,x,y}p_{xy}f(ab|xy)\log_2\left(\frac{f(ab|xy)}{p_\text{NS}(ab|xy)}\right).
\end{equation}
Hence, the optimal no-signaling distribution $\textbf{P}_\text{NS}^*$ is the solution of the optimization
\begin{equation}
    \min_{\textbf{P}_\text{NS} \in P_\text{NS}}D_\text{KL}(\textbf{f}||\textbf{P}_\text{NS}).
\end{equation}
After this estimation, we can find the optimal local realistic distribution $\textbf{P}_\text{LR}^*$ that has the minimum distance from our nearly true probability distribution.
\begin{equation}
    \textbf{P}_\text{LR}^*=\mathop{\arg\min}\limits_{\textbf{P}_\text{LR} \in P_\text{LR}}D_\text{KL}(\textbf{P}^*_\text{NS}||\textbf{P}_\text{LR})=\sum_{a,b,x,y}p_{xy}p^*_\text{NS}(ab|xy)\log_2\left(\frac{p^*_\text{NS}(ab|xy)}{p_\text{LR}(ab|xy)}\right).
\end{equation}
According to Ref.~\cite{Zhang2011Asymptotically}  after $k$ trials but before the $(k + 1)$-st trial the PBR protocol returns a special Bell inequality of the
form
\begin{equation}
\langle R_k(x)\rangle\le1
\end{equation}
with $R_k(x)$ nonnegative. Any such sequence of inequalities can be seen as a  prediction-based radio (PBR) yielding a valid $p$ value computed according to
\begin{equation}
    p^{(PBR)}=\min\left(\left(\prod_{i=1}^nR_{k-1}(x_k)\right)^{-1},1\right)
\end{equation}
The proof can be found in Ref.~\cite{Zhang2011Asymptotically} Sec.III C. 

Once the optimal local realistic distribution $\textbf{P}_\text{LR}^*$ is found, with the local realistic distribution constraint according to Ref.~\cite{Zhang2011Asymptotically}
\begin{equation}
    \sum_{a,b,x,y}\frac{p^*_\text{NS}(ab|xy)}{p_\text{LR}^*(ab|xy)}p_{xy}p_\text{LR}(ab|xy)\le1    
\end{equation}
The test statistics $R(ABXY)\equiv\frac{p^*_\text{NS}(ab|xy)}{p_\text{LR}^*(ab|xy)}$ can be seen as PBR to perform a hypothesis test of local realism. To avoid the i.i.d. (independent and identical distribution) 
assumption in the test, one needs to construct the test statistics $R_i(A_iB_iX_iY_i)$ for the $i$-th
trial before it’s practical outcome, that’s why this method named prediction-based radio method. 
To realize this prediction, we divide the total data into $2.4\times10^7$ trial sized block. For the 
first data block, we set $R(ABXY)=1$. For the following block $i$ we construct the $R_i(A_iB_iX_iY_i)$
based on the frequency distribution of the $i-1$ block. With the PBRs constructed, the $p$ value upper 
bound of the $n$ trials for rejecting local realism is given by:
\begin{equation}
    p_n=\min\left(\left(\prod_{i=1}^nR_i(A_iB_iX_iY_i)\right)^{-1},1\right)    
\end{equation}
where $a_ib_i$ and $x_iy_i$ are the measurement outcomes and setting choices at the $i$-th trial.
The results observed in 6 hours with $n=5.32\times10^9$ show that the $p$ value for obtaining our 
result by the local realism is upper bounded by $10^{-16348}$. This shows an extremely strong 
evidence against local realism provided by our experiment output.

\subsection{Test of no signaling}
We checked whether our experimental data are in agreement with the no-signaling principle by a  traditional hypothesis test with the i.i.d. assumption. In the experiment, there are eight no-signaling 
conditions: Alice obtains 0 or 1 under the setting $x=0$ or $1$ is independent of Bob’s setting choice, 
and Bob obtains $0$ or $1$ under the setting $y=0$ or $1$ is independent of Alice’s setting choice. We don’t 
mention the situation that Alice(Bob)'s outcomes $u$ because if Alice(Bob) obtains $o$ and $e$ independently, then we 
can deduce that she(he) can outcome $u$ freely, too. We construct the test statistics of no-signaling with 
the method named two-proportion Z-test. This method is always used to determine whether two samples are 
drawn from the same population. Generally, if the p-value is larger than 0.05 we can say no evidence of 
violating the no-signaling principle by our experimental data. The p-value that we found is in the following table:

\setlength{\tabcolsep}{10mm}
\begin{table}[ht]
    \centering
    \caption{$p$ values for the hypothesis test of no-signalling}
    \begin{tabular}{c|c|c}
    \hline
    \diagbox{input}{$p$ value}{output} & out = $1$  & out = $0$      \\ \hline
    $x=0$     & 0.3948     & 0.7823         \\
    $x=1$     & 0.7482     & 0.7729         \\
    $y=0$     & 0.4667     & 0.3560         \\
    $y=1$     & 0.0604     & 0.9518         \\ \hline
    \end{tabular}
    \label{tab:nosingal}
    \end{table}
\end{appendix}
%%%%%%%%%%%%%%%%%%%%%%%%%%%%%%%%%%%%%%%%
% choose a style
%\bibliographystyle{ieeetr}
%\bibliographystyle{unsrt}
%%%%%%%%%%%%%%%%%%%%%%%%%%%%%%%%%%%%%%%%

%%%%%%%%%%%%%%%%%%%%%%%%%%%%%%%%%%%%%%%%
% choose a .bib file
\nolinenumbers
%\newpage
\bibliography{bib/Experiment_Hardy.bib}

\begin{thebibliography}{56}%
\makeatletter
\providecommand \@ifxundefined [1]{%
 \@ifx{#1\undefined}
}%
\providecommand \@ifnum [1]{%
 \ifnum #1\expandafter \@firstoftwo
 \else \expandafter \@secondoftwo
 \fi
}%
\providecommand \@ifx [1]{%
 \ifx #1\expandafter \@firstoftwo
 \else \expandafter \@secondoftwo
 \fi
}%
\providecommand \natexlab [1]{#1}%
\providecommand \enquote  [1]{``#1''}%
\providecommand \bibnamefont  [1]{#1}%
\providecommand \bibfnamefont [1]{#1}%
\providecommand \citenamefont [1]{#1}%
\providecommand \href@noop [0]{\@secondoftwo}%
\providecommand \href [0]{\begingroup \@sanitize@url \@href}%
\providecommand \@href[1]{\@@startlink{#1}\@@href}%
\providecommand \@@href[1]{\endgroup#1\@@endlink}%
\providecommand \@sanitize@url [0]{\catcode `\\12\catcode `\$12\catcode
  `\&12\catcode `\#12\catcode `\^12\catcode `\_12\catcode `\%12\relax}%
\providecommand \@@startlink[1]{}%
\providecommand \@@endlink[0]{}%
\providecommand \url  [0]{\begingroup\@sanitize@url \@url }%
\providecommand \@url [1]{\endgroup\@href {#1}{\urlprefix }}%
\providecommand \urlprefix  [0]{URL }%
\providecommand \Eprint [0]{\href }%
\providecommand \doibase [0]{https://doi.org/}%
\providecommand \selectlanguage [0]{\@gobble}%
\providecommand \bibinfo  [0]{\@secondoftwo}%
\providecommand \bibfield  [0]{\@secondoftwo}%
\providecommand \translation [1]{[#1]}%
\providecommand \BibitemOpen [0]{}%
\providecommand \bibitemStop [0]{}%
\providecommand \bibitemNoStop [0]{.\EOS\space}%
\providecommand \EOS [0]{\spacefactor3000\relax}%
\providecommand \BibitemShut  [1]{\csname bibitem#1\endcsname}%
\let\auto@bib@innerbib\@empty
%</preamble>
\bibitem [{\citenamefont {Einstein}\ \emph {et~al.}(1935)\citenamefont
  {Einstein}, \citenamefont {Podolsky},\ and\ \citenamefont
  {Rosen}}]{Einstein1935}%
  \BibitemOpen
  \bibfield  {author} {\bibinfo {author} {\bibfnamefont {A.}~\bibnamefont
  {Einstein}}, \bibinfo {author} {\bibfnamefont {B.}~\bibnamefont {Podolsky}},\
  and\ \bibinfo {author} {\bibfnamefont {N.}~\bibnamefont {Rosen}},\ }\bibfield
   {title} {\bibinfo {title} {Can quantum-mechanical description of physical
  reality be considered complete?},\ }\href
  {https://doi.org/10.1103/PhysRev.47.777} {\bibfield  {journal} {\bibinfo
  {journal} {Phys. Rev.}\ }\textbf {\bibinfo {volume} {47}},\ \bibinfo {pages}
  {777} (\bibinfo {year} {1935})}\BibitemShut {NoStop}%
\bibitem [{\citenamefont {Bohm}(1952)}]{Bohm1952A}%
  \BibitemOpen
  \bibfield  {author} {\bibinfo {author} {\bibfnamefont {D.}~\bibnamefont
  {Bohm}},\ }\bibfield  {title} {\bibinfo {title} {A suggested interpretation
  of the quantum theory in terms of "hidden" variables. {I}},\ }\href
  {https://doi.org/10.1103/PhysRev.85.166} {\bibfield  {journal} {\bibinfo
  {journal} {Phys. Rev.}\ }\textbf {\bibinfo {volume} {85}},\ \bibinfo {pages}
  {166} (\bibinfo {year} {1952})}\BibitemShut {NoStop}%
\bibitem [{\citenamefont {Bell}(1964)}]{bell1964einstein}%
  \BibitemOpen
  \bibfield  {author} {\bibinfo {author} {\bibfnamefont {J.~S.}\ \bibnamefont
  {Bell}},\ }\bibfield  {title} {\bibinfo {title} {On the einstein podolsky
  rosen paradox},\ }\href {https://doi.org/10.1103/PhysicsPhysiqueFizika.1.195}
  {\bibfield  {journal} {\bibinfo  {journal} {Physics}\ }\textbf {\bibinfo
  {volume} {1}},\ \bibinfo {pages} {195} (\bibinfo {year} {1964})}\BibitemShut
  {NoStop}%
\bibitem [{\citenamefont {Clauser}\ \emph {et~al.}(1969)\citenamefont
  {Clauser}, \citenamefont {Horne}, \citenamefont {Shimony},\ and\
  \citenamefont {Holt}}]{clauser1969proposed}%
  \BibitemOpen
  \bibfield  {author} {\bibinfo {author} {\bibfnamefont {J.~F.}\ \bibnamefont
  {Clauser}}, \bibinfo {author} {\bibfnamefont {M.~A.}\ \bibnamefont {Horne}},
  \bibinfo {author} {\bibfnamefont {A.}~\bibnamefont {Shimony}},\ and\ \bibinfo
  {author} {\bibfnamefont {R.~A.}\ \bibnamefont {Holt}},\ }\bibfield  {title}
  {\bibinfo {title} {Proposed experiment to test local hidden-variable
  theories},\ }\href
  {https://doi.org/https://doi.org/10.1103/PhysRevLett.23.880} {\bibfield
  {journal} {\bibinfo  {journal} {Phys. Rev. Lett.}\ }\textbf {\bibinfo
  {volume} {23}},\ \bibinfo {pages} {880} (\bibinfo {year} {1969})}\BibitemShut
  {NoStop}%
\bibitem [{\citenamefont {Brunner}\ \emph {et~al.}(2014)\citenamefont
  {Brunner}, \citenamefont {Cavalcanti}, \citenamefont {Pironio}, \citenamefont
  {Scarani},\ and\ \citenamefont {Wehner}}]{brunner2014bell}%
  \BibitemOpen
  \bibfield  {author} {\bibinfo {author} {\bibfnamefont {N.}~\bibnamefont
  {Brunner}}, \bibinfo {author} {\bibfnamefont {D.}~\bibnamefont {Cavalcanti}},
  \bibinfo {author} {\bibfnamefont {S.}~\bibnamefont {Pironio}}, \bibinfo
  {author} {\bibfnamefont {V.}~\bibnamefont {Scarani}},\ and\ \bibinfo {author}
  {\bibfnamefont {S.}~\bibnamefont {Wehner}},\ }\bibfield  {title} {\bibinfo
  {title} {Bell nonlocality},\ }\href
  {https://doi.org/10.1103/RevModPhys.86.419} {\bibfield  {journal} {\bibinfo
  {journal} {Rev. Mod. Phys.}\ }\textbf {\bibinfo {volume} {86}},\ \bibinfo
  {pages} {419} (\bibinfo {year} {2014})}\BibitemShut {NoStop}%
\bibitem [{\citenamefont {Freedman}\ and\ \citenamefont
  {Clauser}(1972)}]{Freedman1972Experimental}%
  \BibitemOpen
  \bibfield  {author} {\bibinfo {author} {\bibfnamefont {S.~J.}\ \bibnamefont
  {Freedman}}\ and\ \bibinfo {author} {\bibfnamefont {J.~F.}\ \bibnamefont
  {Clauser}},\ }\bibfield  {title} {\bibinfo {title} {Experimental test of
  local hidden-variable theories},\ }\href
  {https://doi.org/10.1103/PhysRevLett.28.938} {\bibfield  {journal} {\bibinfo
  {journal} {Phys. Rev. Lett.}\ }\textbf {\bibinfo {volume} {28}},\ \bibinfo
  {pages} {938} (\bibinfo {year} {1972})}\BibitemShut {NoStop}%
\bibitem [{\citenamefont {Aspect}\ \emph {et~al.}(1982)\citenamefont {Aspect},
  \citenamefont {Grangier},\ and\ \citenamefont
  {Roger}}]{Aspect1982Experimental}%
  \BibitemOpen
  \bibfield  {author} {\bibinfo {author} {\bibfnamefont {A.}~\bibnamefont
  {Aspect}}, \bibinfo {author} {\bibfnamefont {P.}~\bibnamefont {Grangier}},\
  and\ \bibinfo {author} {\bibfnamefont {G.}~\bibnamefont {Roger}},\ }\bibfield
   {title} {\bibinfo {title} {Experimental realization of
  einstein-podolsky-rosen-bohm gedankenexperiment: A new violation of bell's
  inequalities},\ }\href {https://doi.org/10.1103/PhysRevLett.49.91} {\bibfield
   {journal} {\bibinfo  {journal} {Phys. Rev. Lett.}\ }\textbf {\bibinfo
  {volume} {49}},\ \bibinfo {pages} {91} (\bibinfo {year} {1982})}\BibitemShut
  {NoStop}%
\bibitem [{\citenamefont {Hensen}\ \emph {et~al.}(2015)\citenamefont {Hensen},
  \citenamefont {Bernien}, \citenamefont {Dr{\'e}au}, \citenamefont {Reiserer},
  \citenamefont {Kalb}, \citenamefont {Blok}, \citenamefont {Ruitenberg},
  \citenamefont {Vermeulen}, \citenamefont {Schouten}, \citenamefont
  {Abell{\'a}n} \emph {et~al.}}]{hensen2015loophole}%
  \BibitemOpen
  \bibfield  {author} {\bibinfo {author} {\bibfnamefont {B.}~\bibnamefont
  {Hensen}}, \bibinfo {author} {\bibfnamefont {H.}~\bibnamefont {Bernien}},
  \bibinfo {author} {\bibfnamefont {A.~E.}\ \bibnamefont {Dr{\'e}au}}, \bibinfo
  {author} {\bibfnamefont {A.}~\bibnamefont {Reiserer}}, \bibinfo {author}
  {\bibfnamefont {N.}~\bibnamefont {Kalb}}, \bibinfo {author} {\bibfnamefont
  {M.~S.}\ \bibnamefont {Blok}}, \bibinfo {author} {\bibfnamefont
  {J.}~\bibnamefont {Ruitenberg}}, \bibinfo {author} {\bibfnamefont {R.~F.}\
  \bibnamefont {Vermeulen}}, \bibinfo {author} {\bibfnamefont {R.~N.}\
  \bibnamefont {Schouten}}, \bibinfo {author} {\bibfnamefont {C.}~\bibnamefont
  {Abell{\'a}n}}, \emph {et~al.},\ }\bibfield  {title} {\bibinfo {title}
  {Loophole-free {Bell} inequality violation using electron spins separated by
  1.3 kilometres},\ }\href {https://doi.org/10.1038/nature15759} {\bibfield
  {journal} {\bibinfo  {journal} {Nature}\ }\textbf {\bibinfo {volume} {526}},\
  \bibinfo {pages} {682} (\bibinfo {year} {2015})}\BibitemShut {NoStop}%
\bibitem [{\citenamefont {Shalm}\ \emph {et~al.}(2015)\citenamefont {Shalm},
  \citenamefont {Meyer-Scott}, \citenamefont {Christensen}, \citenamefont
  {Bierhorst}, \citenamefont {Wayne}, \citenamefont {Stevens}, \citenamefont
  {Gerrits}, \citenamefont {Glancy}, \citenamefont {Hamel}, \citenamefont
  {Allman} \emph {et~al.}}]{Shalm2015Strong}%
  \BibitemOpen
  \bibfield  {author} {\bibinfo {author} {\bibfnamefont {L.~K.}\ \bibnamefont
  {Shalm}}, \bibinfo {author} {\bibfnamefont {E.}~\bibnamefont {Meyer-Scott}},
  \bibinfo {author} {\bibfnamefont {B.~G.}\ \bibnamefont {Christensen}},
  \bibinfo {author} {\bibfnamefont {P.}~\bibnamefont {Bierhorst}}, \bibinfo
  {author} {\bibfnamefont {M.~A.}\ \bibnamefont {Wayne}}, \bibinfo {author}
  {\bibfnamefont {M.~J.}\ \bibnamefont {Stevens}}, \bibinfo {author}
  {\bibfnamefont {T.}~\bibnamefont {Gerrits}}, \bibinfo {author} {\bibfnamefont
  {S.}~\bibnamefont {Glancy}}, \bibinfo {author} {\bibfnamefont {D.~R.}\
  \bibnamefont {Hamel}}, \bibinfo {author} {\bibfnamefont {M.~S.}\ \bibnamefont
  {Allman}}, \emph {et~al.},\ }\bibfield  {title} {\bibinfo {title} {Strong
  loophole-free test of local realism},\ }\href
  {https://doi.org/10.1103/PhysRevLett.115.250402} {\bibfield  {journal}
  {\bibinfo  {journal} {Phys. Rev. Lett.}\ }\textbf {\bibinfo {volume} {115}},\
  \bibinfo {pages} {250402} (\bibinfo {year} {2015})}\BibitemShut {NoStop}%
\bibitem [{\citenamefont {Giustina}\ \emph {et~al.}(2015)\citenamefont
  {Giustina}, \citenamefont {Versteegh}, \citenamefont {Wengerowsky},
  \citenamefont {Handsteiner}, \citenamefont {Hochrainer}, \citenamefont
  {Phelan}, \citenamefont {Steinlechner}, \citenamefont {Kofler}, \citenamefont
  {Larsson}, \citenamefont {Abell\'an} \emph
  {et~al.}}]{Giustina2015Significant}%
  \BibitemOpen
  \bibfield  {author} {\bibinfo {author} {\bibfnamefont {M.}~\bibnamefont
  {Giustina}}, \bibinfo {author} {\bibfnamefont {M.~A.~M.}\ \bibnamefont
  {Versteegh}}, \bibinfo {author} {\bibfnamefont {S.}~\bibnamefont
  {Wengerowsky}}, \bibinfo {author} {\bibfnamefont {J.}~\bibnamefont
  {Handsteiner}}, \bibinfo {author} {\bibfnamefont {A.}~\bibnamefont
  {Hochrainer}}, \bibinfo {author} {\bibfnamefont {K.}~\bibnamefont {Phelan}},
  \bibinfo {author} {\bibfnamefont {F.}~\bibnamefont {Steinlechner}}, \bibinfo
  {author} {\bibfnamefont {J.}~\bibnamefont {Kofler}}, \bibinfo {author}
  {\bibfnamefont {J.-A.}\ \bibnamefont {Larsson}}, \bibinfo {author}
  {\bibfnamefont {C.}~\bibnamefont {Abell\'an}}, \emph {et~al.},\ }\bibfield
  {title} {\bibinfo {title} {Significant-loophole-free test of {Bell's} theorem
  with entangled photons},\ }\href
  {https://doi.org/10.1103/PhysRevLett.115.250401} {\bibfield  {journal}
  {\bibinfo  {journal} {Phys. Rev. Lett.}\ }\textbf {\bibinfo {volume} {115}},\
  \bibinfo {pages} {250401} (\bibinfo {year} {2015})}\BibitemShut {NoStop}%
\bibitem [{\citenamefont {Li}\ \emph {et~al.}(2018)\citenamefont {Li},
  \citenamefont {Wu}, \citenamefont {Zhang}, \citenamefont {Liu}, \citenamefont
  {Bai}, \citenamefont {Liu}, \citenamefont {Zhang}, \citenamefont {Zhao},
  \citenamefont {Li}, \citenamefont {Wang} \emph {et~al.}}]{Li2018Test}%
  \BibitemOpen
  \bibfield  {author} {\bibinfo {author} {\bibfnamefont {M.-H.}\ \bibnamefont
  {Li}}, \bibinfo {author} {\bibfnamefont {C.}~\bibnamefont {Wu}}, \bibinfo
  {author} {\bibfnamefont {Y.}~\bibnamefont {Zhang}}, \bibinfo {author}
  {\bibfnamefont {W.-Z.}\ \bibnamefont {Liu}}, \bibinfo {author} {\bibfnamefont
  {B.}~\bibnamefont {Bai}}, \bibinfo {author} {\bibfnamefont {Y.}~\bibnamefont
  {Liu}}, \bibinfo {author} {\bibfnamefont {W.}~\bibnamefont {Zhang}}, \bibinfo
  {author} {\bibfnamefont {Q.}~\bibnamefont {Zhao}}, \bibinfo {author}
  {\bibfnamefont {H.}~\bibnamefont {Li}}, \bibinfo {author} {\bibfnamefont
  {Z.}~\bibnamefont {Wang}}, \emph {et~al.},\ }\bibfield  {title} {\bibinfo
  {title} {Test of local realism into the past without detection and locality
  loopholes},\ }\href {https://doi.org/10.1103/PhysRevLett.121.080404}
  {\bibfield  {journal} {\bibinfo  {journal} {Phys. Rev. Lett.}\ }\textbf
  {\bibinfo {volume} {121}},\ \bibinfo {pages} {080404} (\bibinfo {year}
  {2018})}\BibitemShut {NoStop}%
\bibitem [{\citenamefont {Rauch}\ \emph {et~al.}(2018)\citenamefont {Rauch},
  \citenamefont {Handsteiner}, \citenamefont {Hochrainer}, \citenamefont
  {Gallicchio}, \citenamefont {Friedman}, \citenamefont {Leung}, \citenamefont
  {Liu}, \citenamefont {Bulla}, \citenamefont {Ecker}, \citenamefont
  {Steinlechner} \emph {et~al.}}]{Rauch2018Cosmic}%
  \BibitemOpen
  \bibfield  {author} {\bibinfo {author} {\bibfnamefont {D.}~\bibnamefont
  {Rauch}}, \bibinfo {author} {\bibfnamefont {J.}~\bibnamefont {Handsteiner}},
  \bibinfo {author} {\bibfnamefont {A.}~\bibnamefont {Hochrainer}}, \bibinfo
  {author} {\bibfnamefont {J.}~\bibnamefont {Gallicchio}}, \bibinfo {author}
  {\bibfnamefont {A.~S.}\ \bibnamefont {Friedman}}, \bibinfo {author}
  {\bibfnamefont {C.}~\bibnamefont {Leung}}, \bibinfo {author} {\bibfnamefont
  {B.}~\bibnamefont {Liu}}, \bibinfo {author} {\bibfnamefont {L.}~\bibnamefont
  {Bulla}}, \bibinfo {author} {\bibfnamefont {S.}~\bibnamefont {Ecker}},
  \bibinfo {author} {\bibfnamefont {F.}~\bibnamefont {Steinlechner}}, \emph
  {et~al.},\ }\bibfield  {title} {\bibinfo {title} {Cosmic bell test using
  random measurement settings from high-redshift quasars},\ }\href
  {https://doi.org/10.1103/PhysRevLett.121.080403} {\bibfield  {journal}
  {\bibinfo  {journal} {Phys. Rev. Lett.}\ }\textbf {\bibinfo {volume} {121}},\
  \bibinfo {pages} {080403} (\bibinfo {year} {2018})}\BibitemShut {NoStop}%
\bibitem [{\citenamefont {Storz}\ \emph {et~al.}(2023)\citenamefont {Storz},
  \citenamefont {Sch{\"a}r}, \citenamefont {Kulikov}, \citenamefont {Magnard},
  \citenamefont {Kurpiers}, \citenamefont {L{\"u}tolf}, \citenamefont {Walter},
  \citenamefont {Copetudo}, \citenamefont {Reuer}, \citenamefont {Akin} \emph
  {et~al.}}]{storz2023loophole}%
  \BibitemOpen
  \bibfield  {author} {\bibinfo {author} {\bibfnamefont {S.}~\bibnamefont
  {Storz}}, \bibinfo {author} {\bibfnamefont {J.}~\bibnamefont {Sch{\"a}r}},
  \bibinfo {author} {\bibfnamefont {A.}~\bibnamefont {Kulikov}}, \bibinfo
  {author} {\bibfnamefont {P.}~\bibnamefont {Magnard}}, \bibinfo {author}
  {\bibfnamefont {P.}~\bibnamefont {Kurpiers}}, \bibinfo {author}
  {\bibfnamefont {J.}~\bibnamefont {L{\"u}tolf}}, \bibinfo {author}
  {\bibfnamefont {T.}~\bibnamefont {Walter}}, \bibinfo {author} {\bibfnamefont
  {A.}~\bibnamefont {Copetudo}}, \bibinfo {author} {\bibfnamefont
  {K.}~\bibnamefont {Reuer}}, \bibinfo {author} {\bibfnamefont
  {A.}~\bibnamefont {Akin}}, \emph {et~al.},\ }\bibfield  {title} {\bibinfo
  {title} {Loophole-free bell inequality violation with superconducting
  circuits},\ }\href {https://www.nature.com/articles/s41586-023-05885-0}
  {\bibfield  {journal} {\bibinfo  {journal} {Nature}\ }\textbf {\bibinfo
  {volume} {617}},\ \bibinfo {pages} {265} (\bibinfo {year}
  {2023})}\BibitemShut {NoStop}%
\bibitem [{\citenamefont {Mayers}\ and\ \citenamefont
  {Yao}(1998)}]{mayers1998quantum}%
  \BibitemOpen
  \bibfield  {author} {\bibinfo {author} {\bibfnamefont {D.}~\bibnamefont
  {Mayers}}\ and\ \bibinfo {author} {\bibfnamefont {A.}~\bibnamefont {Yao}},\
  }\bibfield  {title} {\bibinfo {title} {Quantum cryptography with imperfect
  apparatus},\ }in\ \href@noop {} {\emph {\bibinfo {booktitle} {Proceedings
  39th Annual Symposium on Foundations of Computer Science (Cat. No.
  98CB36280)}}}\ (\bibinfo {organization} {IEEE},\ \bibinfo {year} {1998})\
  pp.\ \bibinfo {pages} {503--509}\BibitemShut {NoStop}%
\bibitem [{\citenamefont {Barrett}\ \emph {et~al.}(2005)\citenamefont
  {Barrett}, \citenamefont {Hardy},\ and\ \citenamefont
  {Kent}}]{Barrett2005no}%
  \BibitemOpen
  \bibfield  {author} {\bibinfo {author} {\bibfnamefont {J.}~\bibnamefont
  {Barrett}}, \bibinfo {author} {\bibfnamefont {L.}~\bibnamefont {Hardy}},\
  and\ \bibinfo {author} {\bibfnamefont {A.}~\bibnamefont {Kent}},\ }\bibfield
  {title} {\bibinfo {title} {No signaling and quantum key distribution},\
  }\href {https://doi.org/10.1103/PhysRevLett.95.010503} {\bibfield  {journal}
  {\bibinfo  {journal} {Phys. Rev. Lett.}\ }\textbf {\bibinfo {volume} {95}},\
  \bibinfo {pages} {010503} (\bibinfo {year} {2005})}\BibitemShut {NoStop}%
\bibitem [{\citenamefont {Ac\'{\i}n}\ \emph {et~al.}(2006)\citenamefont
  {Ac\'{\i}n}, \citenamefont {Gisin},\ and\ \citenamefont
  {Masanes}}]{Acin2006from}%
  \BibitemOpen
  \bibfield  {author} {\bibinfo {author} {\bibfnamefont {A.}~\bibnamefont
  {Ac\'{\i}n}}, \bibinfo {author} {\bibfnamefont {N.}~\bibnamefont {Gisin}},\
  and\ \bibinfo {author} {\bibfnamefont {L.}~\bibnamefont {Masanes}},\
  }\bibfield  {title} {\bibinfo {title} {From bell's theorem to secure quantum
  key distribution},\ }\href {https://doi.org/10.1103/PhysRevLett.97.120405}
  {\bibfield  {journal} {\bibinfo  {journal} {Phys. Rev. Lett.}\ }\textbf
  {\bibinfo {volume} {97}},\ \bibinfo {pages} {120405} (\bibinfo {year}
  {2006})}\BibitemShut {NoStop}%
\bibitem [{\citenamefont {Ac\'{\i}n}\ \emph {et~al.}(2007)\citenamefont
  {Ac\'{\i}n}, \citenamefont {Brunner}, \citenamefont {Gisin}, \citenamefont
  {Massar}, \citenamefont {Pironio},\ and\ \citenamefont
  {Scarani}}]{Acin2007device}%
  \BibitemOpen
  \bibfield  {author} {\bibinfo {author} {\bibfnamefont {A.}~\bibnamefont
  {Ac\'{\i}n}}, \bibinfo {author} {\bibfnamefont {N.}~\bibnamefont {Brunner}},
  \bibinfo {author} {\bibfnamefont {N.}~\bibnamefont {Gisin}}, \bibinfo
  {author} {\bibfnamefont {S.}~\bibnamefont {Massar}}, \bibinfo {author}
  {\bibfnamefont {S.}~\bibnamefont {Pironio}},\ and\ \bibinfo {author}
  {\bibfnamefont {V.}~\bibnamefont {Scarani}},\ }\bibfield  {title} {\bibinfo
  {title} {Device-independent security of quantum cryptography against
  collective attacks},\ }\href {https://doi.org/10.1103/PhysRevLett.98.230501}
  {\bibfield  {journal} {\bibinfo  {journal} {Phys. Rev. Lett.}\ }\textbf
  {\bibinfo {volume} {98}},\ \bibinfo {pages} {230501} (\bibinfo {year}
  {2007})}\BibitemShut {NoStop}%
\bibitem [{\citenamefont {Colbeck}(2007)}]{colbeck2007quantum}%
  \BibitemOpen
  \bibfield  {author} {\bibinfo {author} {\bibfnamefont {R.}~\bibnamefont
  {Colbeck}},\ }\bibfield  {title} {\bibinfo {title} {Quantum and relativistic
  protocols for secure multi-party computation},\ }\href
  {https://arxiv.org/abs/0911.3814} {\bibfield  {journal} {\bibinfo  {journal}
  {Ph.D. thesis, Univ. of Cambridge}\ } (\bibinfo {year} {2007})}\BibitemShut
  {NoStop}%
\bibitem [{\citenamefont {Pironio}\ \emph {et~al.}(2010)\citenamefont
  {Pironio}, \citenamefont {Ac{\'\i}n}, \citenamefont {Massar}, \citenamefont
  {de~La~Giroday}, \citenamefont {Matsukevich}, \citenamefont {Maunz},
  \citenamefont {Olmschenk}, \citenamefont {Hayes}, \citenamefont {Luo},
  \citenamefont {Manning} \emph {et~al.}}]{pironio2010random}%
  \BibitemOpen
  \bibfield  {author} {\bibinfo {author} {\bibfnamefont {S.}~\bibnamefont
  {Pironio}}, \bibinfo {author} {\bibfnamefont {A.}~\bibnamefont {Ac{\'\i}n}},
  \bibinfo {author} {\bibfnamefont {S.}~\bibnamefont {Massar}}, \bibinfo
  {author} {\bibfnamefont {A.~B.}\ \bibnamefont {de~La~Giroday}}, \bibinfo
  {author} {\bibfnamefont {D.~N.}\ \bibnamefont {Matsukevich}}, \bibinfo
  {author} {\bibfnamefont {P.}~\bibnamefont {Maunz}}, \bibinfo {author}
  {\bibfnamefont {S.}~\bibnamefont {Olmschenk}}, \bibinfo {author}
  {\bibfnamefont {D.}~\bibnamefont {Hayes}}, \bibinfo {author} {\bibfnamefont
  {L.}~\bibnamefont {Luo}}, \bibinfo {author} {\bibfnamefont {T.~A.}\
  \bibnamefont {Manning}}, \emph {et~al.},\ }\bibfield  {title} {\bibinfo
  {title} {Random numbers certified by bell’s theorem},\ }\href
  {https://www.nature.com/articles/nature09008} {\bibfield  {journal} {\bibinfo
   {journal} {Nature}\ }\textbf {\bibinfo {volume} {464}},\ \bibinfo {pages}
  {1021} (\bibinfo {year} {2010})}\BibitemShut {NoStop}%
\bibitem [{\citenamefont {Colbeck}\ and\ \citenamefont
  {Renner}(2012)}]{colbeck2012free}%
  \BibitemOpen
  \bibfield  {author} {\bibinfo {author} {\bibfnamefont {R.}~\bibnamefont
  {Colbeck}}\ and\ \bibinfo {author} {\bibfnamefont {R.}~\bibnamefont
  {Renner}},\ }\bibfield  {title} {\bibinfo {title} {Free randomness can be
  amplified},\ }\href {https://doi.org/10.1038/nphys2300} {\bibfield  {journal}
  {\bibinfo  {journal} {Nat. Phys.}\ }\textbf {\bibinfo {volume} {8}},\
  \bibinfo {pages} {450} (\bibinfo {year} {2012})}\BibitemShut {NoStop}%
\bibitem [{\citenamefont {Bierhorst}\ \emph {et~al.}(2018)\citenamefont
  {Bierhorst}, \citenamefont {Knill}, \citenamefont {Glancy}, \citenamefont
  {Zhang}, \citenamefont {Mink}, \citenamefont {Jordan}, \citenamefont
  {Rommal}, \citenamefont {Liu}, \citenamefont {Christensen}, \citenamefont
  {Nam} \emph {et~al.}}]{bierhorst2018experimentally}%
  \BibitemOpen
  \bibfield  {author} {\bibinfo {author} {\bibfnamefont {P.}~\bibnamefont
  {Bierhorst}}, \bibinfo {author} {\bibfnamefont {E.}~\bibnamefont {Knill}},
  \bibinfo {author} {\bibfnamefont {S.}~\bibnamefont {Glancy}}, \bibinfo
  {author} {\bibfnamefont {Y.}~\bibnamefont {Zhang}}, \bibinfo {author}
  {\bibfnamefont {A.}~\bibnamefont {Mink}}, \bibinfo {author} {\bibfnamefont
  {S.}~\bibnamefont {Jordan}}, \bibinfo {author} {\bibfnamefont
  {A.}~\bibnamefont {Rommal}}, \bibinfo {author} {\bibfnamefont {Y.-K.}\
  \bibnamefont {Liu}}, \bibinfo {author} {\bibfnamefont {B.}~\bibnamefont
  {Christensen}}, \bibinfo {author} {\bibfnamefont {S.~W.}\ \bibnamefont
  {Nam}}, \emph {et~al.},\ }\bibfield  {title} {\bibinfo {title}
  {Experimentally generated randomness certified by the impossibility of
  superluminal signals},\ }\href
  {https://www.nature.com/articles/s41586-018-0019-0} {\bibfield  {journal}
  {\bibinfo  {journal} {Nature}\ }\textbf {\bibinfo {volume} {556}},\ \bibinfo
  {pages} {223} (\bibinfo {year} {2018})}\BibitemShut {NoStop}%
\bibitem [{\citenamefont {Liu}\ \emph {et~al.}(2018)\citenamefont {Liu},
  \citenamefont {Zhao}, \citenamefont {Li}, \citenamefont {Guan}, \citenamefont
  {Zhang}, \citenamefont {Bai}, \citenamefont {Zhang}, \citenamefont {Liu},
  \citenamefont {Wu}, \citenamefont {Yuan} \emph {et~al.}}]{liu2018device}%
  \BibitemOpen
  \bibfield  {author} {\bibinfo {author} {\bibfnamefont {Y.}~\bibnamefont
  {Liu}}, \bibinfo {author} {\bibfnamefont {Q.}~\bibnamefont {Zhao}}, \bibinfo
  {author} {\bibfnamefont {M.-H.}\ \bibnamefont {Li}}, \bibinfo {author}
  {\bibfnamefont {J.-Y.}\ \bibnamefont {Guan}}, \bibinfo {author}
  {\bibfnamefont {Y.}~\bibnamefont {Zhang}}, \bibinfo {author} {\bibfnamefont
  {B.}~\bibnamefont {Bai}}, \bibinfo {author} {\bibfnamefont {W.}~\bibnamefont
  {Zhang}}, \bibinfo {author} {\bibfnamefont {W.-Z.}\ \bibnamefont {Liu}},
  \bibinfo {author} {\bibfnamefont {C.}~\bibnamefont {Wu}}, \bibinfo {author}
  {\bibfnamefont {X.}~\bibnamefont {Yuan}}, \emph {et~al.},\ }\bibfield
  {title} {\bibinfo {title} {Device-independent quantum random-number
  generation},\ }\href
  {https://doi.org/https://doi.org/10.1038/s41586-018-0559-3} {\bibfield
  {journal} {\bibinfo  {journal} {Nature}\ }\textbf {\bibinfo {volume} {562}},\
  \bibinfo {pages} {548} (\bibinfo {year} {2018})}\BibitemShut {NoStop}%
\bibitem [{\citenamefont {Zhang}\ \emph {et~al.}(2020)\citenamefont {Zhang},
  \citenamefont {Shalm}, \citenamefont {Bienfang}, \citenamefont {Stevens},
  \citenamefont {Mazurek}, \citenamefont {Nam}, \citenamefont {Abell\'an},
  \citenamefont {Amaya}, \citenamefont {Mitchell}, \citenamefont {Fu} \emph
  {et~al.}}]{Zhang2020Experimental}%
  \BibitemOpen
  \bibfield  {author} {\bibinfo {author} {\bibfnamefont {Y.}~\bibnamefont
  {Zhang}}, \bibinfo {author} {\bibfnamefont {L.~K.}\ \bibnamefont {Shalm}},
  \bibinfo {author} {\bibfnamefont {J.~C.}\ \bibnamefont {Bienfang}}, \bibinfo
  {author} {\bibfnamefont {M.~J.}\ \bibnamefont {Stevens}}, \bibinfo {author}
  {\bibfnamefont {M.~D.}\ \bibnamefont {Mazurek}}, \bibinfo {author}
  {\bibfnamefont {S.~W.}\ \bibnamefont {Nam}}, \bibinfo {author} {\bibfnamefont
  {C.}~\bibnamefont {Abell\'an}}, \bibinfo {author} {\bibfnamefont
  {W.}~\bibnamefont {Amaya}}, \bibinfo {author} {\bibfnamefont {M.~W.}\
  \bibnamefont {Mitchell}}, \bibinfo {author} {\bibfnamefont {H.}~\bibnamefont
  {Fu}}, \emph {et~al.},\ }\bibfield  {title} {\bibinfo {title} {Experimental
  low-latency device-independent quantum randomness},\ }\href
  {https://doi.org/10.1103/PhysRevLett.124.010505} {\bibfield  {journal}
  {\bibinfo  {journal} {Phys. Rev. Lett.}\ }\textbf {\bibinfo {volume} {124}},\
  \bibinfo {pages} {010505} (\bibinfo {year} {2020})}\BibitemShut {NoStop}%
\bibitem [{\citenamefont {Shalm}\ \emph {et~al.}(2021)\citenamefont {Shalm},
  \citenamefont {Zhang}, \citenamefont {Bienfang}, \citenamefont {Schlager},
  \citenamefont {Stevens}, \citenamefont {Mazurek}, \citenamefont
  {Abell{\'a}n}, \citenamefont {Amaya}, \citenamefont {Mitchell}, \citenamefont
  {Alhejji} \emph {et~al.}}]{shalm2021device}%
  \BibitemOpen
  \bibfield  {author} {\bibinfo {author} {\bibfnamefont {L.~K.}\ \bibnamefont
  {Shalm}}, \bibinfo {author} {\bibfnamefont {Y.}~\bibnamefont {Zhang}},
  \bibinfo {author} {\bibfnamefont {J.~C.}\ \bibnamefont {Bienfang}}, \bibinfo
  {author} {\bibfnamefont {C.}~\bibnamefont {Schlager}}, \bibinfo {author}
  {\bibfnamefont {M.~J.}\ \bibnamefont {Stevens}}, \bibinfo {author}
  {\bibfnamefont {M.~D.}\ \bibnamefont {Mazurek}}, \bibinfo {author}
  {\bibfnamefont {C.}~\bibnamefont {Abell{\'a}n}}, \bibinfo {author}
  {\bibfnamefont {W.}~\bibnamefont {Amaya}}, \bibinfo {author} {\bibfnamefont
  {M.~W.}\ \bibnamefont {Mitchell}}, \bibinfo {author} {\bibfnamefont {M.~A.}\
  \bibnamefont {Alhejji}}, \emph {et~al.},\ }\bibfield  {title} {\bibinfo
  {title} {Device-independent randomness expansion with entangled photons},\
  }\href {https://www.nature.com/articles/s41567-020-01153-4} {\bibfield
  {journal} {\bibinfo  {journal} {Nat. Phys.}\ }\textbf {\bibinfo {volume}
  {17}},\ \bibinfo {pages} {452} (\bibinfo {year} {2021})}\BibitemShut
  {NoStop}%
\bibitem [{\citenamefont {Li}\ \emph {et~al.}(2021)\citenamefont {Li},
  \citenamefont {Zhang}, \citenamefont {Liu}, \citenamefont {Zhao},
  \citenamefont {Bai}, \citenamefont {Liu}, \citenamefont {Zhao}, \citenamefont
  {Peng}, \citenamefont {Zhang}, \citenamefont {Zhang} \emph
  {et~al.}}]{Li2021Experimental}%
  \BibitemOpen
  \bibfield  {author} {\bibinfo {author} {\bibfnamefont {M.-H.}\ \bibnamefont
  {Li}}, \bibinfo {author} {\bibfnamefont {X.}~\bibnamefont {Zhang}}, \bibinfo
  {author} {\bibfnamefont {W.-Z.}\ \bibnamefont {Liu}}, \bibinfo {author}
  {\bibfnamefont {S.-R.}\ \bibnamefont {Zhao}}, \bibinfo {author}
  {\bibfnamefont {B.}~\bibnamefont {Bai}}, \bibinfo {author} {\bibfnamefont
  {Y.}~\bibnamefont {Liu}}, \bibinfo {author} {\bibfnamefont {Q.}~\bibnamefont
  {Zhao}}, \bibinfo {author} {\bibfnamefont {Y.}~\bibnamefont {Peng}}, \bibinfo
  {author} {\bibfnamefont {J.}~\bibnamefont {Zhang}}, \bibinfo {author}
  {\bibfnamefont {Y.}~\bibnamefont {Zhang}}, \emph {et~al.},\ }\bibfield
  {title} {\bibinfo {title} {Experimental realization of device-independent
  quantum randomness expansion},\ }\href
  {https://doi.org/10.1103/PhysRevLett.126.050503} {\bibfield  {journal}
  {\bibinfo  {journal} {Phys. Rev. Lett.}\ }\textbf {\bibinfo {volume} {126}},\
  \bibinfo {pages} {050503} (\bibinfo {year} {2021})}\BibitemShut {NoStop}%
\bibitem [{\citenamefont {Liu}\ \emph {et~al.}(2022)\citenamefont {Liu},
  \citenamefont {Zhang}, \citenamefont {Zhen}, \citenamefont {Li},
  \citenamefont {Liu}, \citenamefont {Fan}, \citenamefont {Xu}, \citenamefont
  {Zhang},\ and\ \citenamefont {Pan}}]{Liu2022Toward}%
  \BibitemOpen
  \bibfield  {author} {\bibinfo {author} {\bibfnamefont {W.-Z.}\ \bibnamefont
  {Liu}}, \bibinfo {author} {\bibfnamefont {Y.-Z.}\ \bibnamefont {Zhang}},
  \bibinfo {author} {\bibfnamefont {Y.-Z.}\ \bibnamefont {Zhen}}, \bibinfo
  {author} {\bibfnamefont {M.-H.}\ \bibnamefont {Li}}, \bibinfo {author}
  {\bibfnamefont {Y.}~\bibnamefont {Liu}}, \bibinfo {author} {\bibfnamefont
  {J.}~\bibnamefont {Fan}}, \bibinfo {author} {\bibfnamefont {F.}~\bibnamefont
  {Xu}}, \bibinfo {author} {\bibfnamefont {Q.}~\bibnamefont {Zhang}},\ and\
  \bibinfo {author} {\bibfnamefont {J.-W.}\ \bibnamefont {Pan}},\ }\bibfield
  {title} {\bibinfo {title} {Toward a photonic demonstration of
  device-independent quantum key distribution},\ }\href
  {https://doi.org/10.1103/PhysRevLett.129.050502} {\bibfield  {journal}
  {\bibinfo  {journal} {Phys. Rev. Lett.}\ }\textbf {\bibinfo {volume} {129}},\
  \bibinfo {pages} {050502} (\bibinfo {year} {2022})}\BibitemShut {NoStop}%
\bibitem [{\citenamefont {Xu}\ \emph {et~al.}(2022)\citenamefont {Xu},
  \citenamefont {Zhang}, \citenamefont {Zhang},\ and\ \citenamefont
  {Pan}}]{Xu2022Device}%
  \BibitemOpen
  \bibfield  {author} {\bibinfo {author} {\bibfnamefont {F.}~\bibnamefont
  {Xu}}, \bibinfo {author} {\bibfnamefont {Y.-Z.}\ \bibnamefont {Zhang}},
  \bibinfo {author} {\bibfnamefont {Q.}~\bibnamefont {Zhang}},\ and\ \bibinfo
  {author} {\bibfnamefont {J.-W.}\ \bibnamefont {Pan}},\ }\bibfield  {title}
  {\bibinfo {title} {Device-independent quantum key distribution with random
  postselection},\ }\href {https://doi.org/10.1103/PhysRevLett.128.110506}
  {\bibfield  {journal} {\bibinfo  {journal} {Phys. Rev. Lett.}\ }\textbf
  {\bibinfo {volume} {128}},\ \bibinfo {pages} {110506} (\bibinfo {year}
  {2022})}\BibitemShut {NoStop}%
\bibitem [{\citenamefont {Nadlinger}\ \emph {et~al.}(2022)\citenamefont
  {Nadlinger}, \citenamefont {Drmota}, \citenamefont {Nichol}, \citenamefont
  {Araneda}, \citenamefont {Main}, \citenamefont {Srinivas}, \citenamefont
  {Lucas}, \citenamefont {Ballance}, \citenamefont {Ivanov}, \citenamefont
  {Tan} \emph {et~al.}}]{nadlinger2022experimental}%
  \BibitemOpen
  \bibfield  {author} {\bibinfo {author} {\bibfnamefont {D.~P.}\ \bibnamefont
  {Nadlinger}}, \bibinfo {author} {\bibfnamefont {P.}~\bibnamefont {Drmota}},
  \bibinfo {author} {\bibfnamefont {B.~C.}\ \bibnamefont {Nichol}}, \bibinfo
  {author} {\bibfnamefont {G.}~\bibnamefont {Araneda}}, \bibinfo {author}
  {\bibfnamefont {D.}~\bibnamefont {Main}}, \bibinfo {author} {\bibfnamefont
  {R.}~\bibnamefont {Srinivas}}, \bibinfo {author} {\bibfnamefont {D.~M.}\
  \bibnamefont {Lucas}}, \bibinfo {author} {\bibfnamefont {C.~J.}\ \bibnamefont
  {Ballance}}, \bibinfo {author} {\bibfnamefont {K.}~\bibnamefont {Ivanov}},
  \bibinfo {author} {\bibfnamefont {E.-Z.}\ \bibnamefont {Tan}}, \emph
  {et~al.},\ }\bibfield  {title} {\bibinfo {title} {Experimental quantum key
  distribution certified by {Bell's} theorem},\ }\href
  {https://doi.org/https://doi.org/10.1038/s41586-022-04941-5} {\bibfield
  {journal} {\bibinfo  {journal} {Nature}\ }\textbf {\bibinfo {volume} {607}},\
  \bibinfo {pages} {682} (\bibinfo {year} {2022})}\BibitemShut {NoStop}%
\bibitem [{\citenamefont {Zhang}\ \emph {et~al.}(2022)\citenamefont {Zhang},
  \citenamefont {van Leent}, \citenamefont {Redeker}, \citenamefont {Garthoff},
  \citenamefont {Schwonnek}, \citenamefont {Fertig}, \citenamefont {Eppelt},
  \citenamefont {Rosenfeld}, \citenamefont {Scarani}, \citenamefont {Lim} \emph
  {et~al.}}]{zhang2022device}%
  \BibitemOpen
  \bibfield  {author} {\bibinfo {author} {\bibfnamefont {W.}~\bibnamefont
  {Zhang}}, \bibinfo {author} {\bibfnamefont {T.}~\bibnamefont {van Leent}},
  \bibinfo {author} {\bibfnamefont {K.}~\bibnamefont {Redeker}}, \bibinfo
  {author} {\bibfnamefont {R.}~\bibnamefont {Garthoff}}, \bibinfo {author}
  {\bibfnamefont {R.}~\bibnamefont {Schwonnek}}, \bibinfo {author}
  {\bibfnamefont {F.}~\bibnamefont {Fertig}}, \bibinfo {author} {\bibfnamefont
  {S.}~\bibnamefont {Eppelt}}, \bibinfo {author} {\bibfnamefont
  {W.}~\bibnamefont {Rosenfeld}}, \bibinfo {author} {\bibfnamefont
  {V.}~\bibnamefont {Scarani}}, \bibinfo {author} {\bibfnamefont {C.~C.-W.}\
  \bibnamefont {Lim}}, \emph {et~al.},\ }\bibfield  {title} {\bibinfo {title}
  {A device-independent quantum key distribution system for distant users},\
  }\href {https://doi.org/10.1038/s41586-022-04891-y} {\bibfield  {journal}
  {\bibinfo  {journal} {Nature}\ }\textbf {\bibinfo {volume} {607}},\ \bibinfo
  {pages} {687} (\bibinfo {year} {2022})}\BibitemShut {NoStop}%
\bibitem [{\citenamefont {Li}\ \emph {et~al.}(2023)\citenamefont {Li},
  \citenamefont {Zhang}, \citenamefont {Zhang}, \citenamefont {Yang},
  \citenamefont {Han}, \citenamefont {Cheng}, \citenamefont {Cui},
  \citenamefont {Liu}, \citenamefont {Li}, \citenamefont {Liu} \emph
  {et~al.}}]{li2023device}%
  \BibitemOpen
  \bibfield  {author} {\bibinfo {author} {\bibfnamefont {C.-L.}\ \bibnamefont
  {Li}}, \bibinfo {author} {\bibfnamefont {K.-Y.}\ \bibnamefont {Zhang}},
  \bibinfo {author} {\bibfnamefont {X.}~\bibnamefont {Zhang}}, \bibinfo
  {author} {\bibfnamefont {K.-X.}\ \bibnamefont {Yang}}, \bibinfo {author}
  {\bibfnamefont {Y.}~\bibnamefont {Han}}, \bibinfo {author} {\bibfnamefont
  {S.-Y.}\ \bibnamefont {Cheng}}, \bibinfo {author} {\bibfnamefont
  {H.}~\bibnamefont {Cui}}, \bibinfo {author} {\bibfnamefont {W.-Z.}\
  \bibnamefont {Liu}}, \bibinfo {author} {\bibfnamefont {M.-H.}\ \bibnamefont
  {Li}}, \bibinfo {author} {\bibfnamefont {Y.}~\bibnamefont {Liu}}, \emph
  {et~al.},\ }\bibfield  {title} {\bibinfo {title} {Device-independent quantum
  randomness--enhanced zero-knowledge proof},\ }\href
  {https://www.pnas.org/doi/abs/10.1073/pnas.2205463120} {\bibfield  {journal}
  {\bibinfo  {journal} {Proc. Natl. Acad. Sci.}\ }\textbf {\bibinfo {volume}
  {120}},\ \bibinfo {pages} {e2205463120} (\bibinfo {year} {2023})}\BibitemShut
  {NoStop}%
\bibitem [{\citenamefont {Zapatero}\ \emph {et~al.}(2023)\citenamefont
  {Zapatero}, \citenamefont {van Leent}, \citenamefont {Arnon-Friedman},
  \citenamefont {Liu}, \citenamefont {Zhang}, \citenamefont {Weinfurter},\ and\
  \citenamefont {Curty}}]{zapatero2023advances}%
  \BibitemOpen
  \bibfield  {author} {\bibinfo {author} {\bibfnamefont {V.}~\bibnamefont
  {Zapatero}}, \bibinfo {author} {\bibfnamefont {T.}~\bibnamefont {van Leent}},
  \bibinfo {author} {\bibfnamefont {R.}~\bibnamefont {Arnon-Friedman}},
  \bibinfo {author} {\bibfnamefont {W.-Z.}\ \bibnamefont {Liu}}, \bibinfo
  {author} {\bibfnamefont {Q.}~\bibnamefont {Zhang}}, \bibinfo {author}
  {\bibfnamefont {H.}~\bibnamefont {Weinfurter}},\ and\ \bibinfo {author}
  {\bibfnamefont {M.}~\bibnamefont {Curty}},\ }\bibfield  {title} {\bibinfo
  {title} {Advances in device-independent quantum key distribution},\ }\href
  {https://www.nature.com/articles/s41534-023-00684-x} {\bibfield  {journal}
  {\bibinfo  {journal} {npj Quantum Inf.}\ }\textbf {\bibinfo {volume} {9}},\
  \bibinfo {pages} {10} (\bibinfo {year} {2023})}\BibitemShut {NoStop}%
\bibitem [{\citenamefont {{\v{S}}upi{\'c}}\ \emph {et~al.}(2023)\citenamefont
  {{\v{S}}upi{\'c}}, \citenamefont {Bowles}, \citenamefont {Renou},
  \citenamefont {Ac{\'\i}n},\ and\ \citenamefont {Hoban}}]{vsupic2023quantum}%
  \BibitemOpen
  \bibfield  {author} {\bibinfo {author} {\bibfnamefont {I.}~\bibnamefont
  {{\v{S}}upi{\'c}}}, \bibinfo {author} {\bibfnamefont {J.}~\bibnamefont
  {Bowles}}, \bibinfo {author} {\bibfnamefont {M.-O.}\ \bibnamefont {Renou}},
  \bibinfo {author} {\bibfnamefont {A.}~\bibnamefont {Ac{\'\i}n}},\ and\
  \bibinfo {author} {\bibfnamefont {M.~J.}\ \bibnamefont {Hoban}},\ }\bibfield
  {title} {\bibinfo {title} {Quantum networks self-test all entangled states},\
  }\href {https://www.nature.com/articles/s41567-023-01945-4} {\bibfield
  {journal} {\bibinfo  {journal} {Nat. Phys.}\ }\textbf {\bibinfo {volume}
  {19}},\ \bibinfo {pages} {670} (\bibinfo {year} {2023})}\BibitemShut
  {NoStop}%
\bibitem [{\citenamefont {Greenberger}\ \emph {et~al.}(1989)\citenamefont
  {Greenberger}, \citenamefont {Horne},\ and\ \citenamefont
  {Zeilinger}}]{greenberger1989going}%
  \BibitemOpen
  \bibfield  {author} {\bibinfo {author} {\bibfnamefont {D.~M.}\ \bibnamefont
  {Greenberger}}, \bibinfo {author} {\bibfnamefont {M.~A.}\ \bibnamefont
  {Horne}},\ and\ \bibinfo {author} {\bibfnamefont {A.}~\bibnamefont
  {Zeilinger}},\ }\bibinfo {title} {{Going Beyond Bell's Theorem}},\ in\ \href
  {https://doi.org/10.1007/978-94-017-0849-4_10} {\emph {\bibinfo {booktitle}
  {Bell's Theorem, Quantum Theory and Conceptions of the Universe}}},\ \bibinfo
  {editor} {edited by\ \bibinfo {editor} {\bibfnamefont {M.}~\bibnamefont
  {Kafatos}}}\ (\bibinfo  {publisher} {Springer Netherlands},\ \bibinfo
  {address} {Dordrecht},\ \bibinfo {year} {1989})\ pp.\ \bibinfo {pages}
  {69--72}\BibitemShut {NoStop}%
\bibitem [{\citenamefont {Greenberger}\ \emph {et~al.}(1990)\citenamefont
  {Greenberger}, \citenamefont {Horne}, \citenamefont {Shimony},\ and\
  \citenamefont {Zeilinger}}]{greenberger1990bell}%
  \BibitemOpen
  \bibfield  {author} {\bibinfo {author} {\bibfnamefont {D.~M.}\ \bibnamefont
  {Greenberger}}, \bibinfo {author} {\bibfnamefont {M.~A.}\ \bibnamefont
  {Horne}}, \bibinfo {author} {\bibfnamefont {A.}~\bibnamefont {Shimony}},\
  and\ \bibinfo {author} {\bibfnamefont {A.}~\bibnamefont {Zeilinger}},\
  }\bibfield  {title} {\bibinfo {title} {Bell’s theorem without
  inequalities},\ }\href {https://doi.org/10.1119/1.16243} {\bibfield
  {journal} {\bibinfo  {journal} {Am. J. Phys.}\ }\textbf {\bibinfo {volume}
  {58}},\ \bibinfo {pages} {1131} (\bibinfo {year} {1990})}\BibitemShut
  {NoStop}%
\bibitem [{\citenamefont {Hardy}(1993)}]{hardy1993nonlocality}%
  \BibitemOpen
  \bibfield  {author} {\bibinfo {author} {\bibfnamefont {L.}~\bibnamefont
  {Hardy}},\ }\bibfield  {title} {\bibinfo {title} {Nonlocality for two
  particles without inequalities for almost all entangled states},\ }\href
  {https://doi.org/10.1103/PhysRevLett.71.1665} {\bibfield  {journal} {\bibinfo
   {journal} {Phys. Rev. Lett.}\ }\textbf {\bibinfo {volume} {71}},\ \bibinfo
  {pages} {1665} (\bibinfo {year} {1993})}\BibitemShut {NoStop}%
\bibitem [{\citenamefont {Mermin}(1994)}]{mermin1994quantum}%
  \BibitemOpen
  \bibfield  {author} {\bibinfo {author} {\bibfnamefont {N.~D.}\ \bibnamefont
  {Mermin}},\ }\bibfield  {title} {\bibinfo {title} {Quantum mysteries
  refined},\ }\href {https://doi.org/10.1119/1.17733} {\bibfield  {journal}
  {\bibinfo  {journal} {Am. J. Phys.}\ }\textbf {\bibinfo {volume} {62}},\
  \bibinfo {pages} {880} (\bibinfo {year} {1994})}\BibitemShut {NoStop}%
\bibitem [{\citenamefont {Mukherjee}\ \emph {et~al.}(2015)\citenamefont
  {Mukherjee}, \citenamefont {Roy}, \citenamefont {Bhattacharya}, \citenamefont
  {Das}, \citenamefont {Gazi},\ and\ \citenamefont
  {Banik}}]{Mukherjee2015Hardy}%
  \BibitemOpen
  \bibfield  {author} {\bibinfo {author} {\bibfnamefont {A.}~\bibnamefont
  {Mukherjee}}, \bibinfo {author} {\bibfnamefont {A.}~\bibnamefont {Roy}},
  \bibinfo {author} {\bibfnamefont {S.~S.}\ \bibnamefont {Bhattacharya}},
  \bibinfo {author} {\bibfnamefont {S.}~\bibnamefont {Das}}, \bibinfo {author}
  {\bibfnamefont {M.~R.}\ \bibnamefont {Gazi}},\ and\ \bibinfo {author}
  {\bibfnamefont {M.}~\bibnamefont {Banik}},\ }\bibfield  {title} {\bibinfo
  {title} {Hardy's test as a device-independent dimension witness},\ }\href
  {https://doi.org/10.1103/PhysRevA.92.022302} {\bibfield  {journal} {\bibinfo
  {journal} {Phys. Rev. A}\ }\textbf {\bibinfo {volume} {92}},\ \bibinfo
  {pages} {022302} (\bibinfo {year} {2015})}\BibitemShut {NoStop}%
\bibitem [{\citenamefont {Li}\ \emph {et~al.}(2015)\citenamefont {Li},
  \citenamefont {Paw\l{}owski}, \citenamefont {Rahaman}, \citenamefont {Guo},\
  and\ \citenamefont {Han}}]{Li2015Device}%
  \BibitemOpen
  \bibfield  {author} {\bibinfo {author} {\bibfnamefont {H.-W.}\ \bibnamefont
  {Li}}, \bibinfo {author} {\bibfnamefont {M.}~\bibnamefont {Paw\l{}owski}},
  \bibinfo {author} {\bibfnamefont {R.}~\bibnamefont {Rahaman}}, \bibinfo
  {author} {\bibfnamefont {G.-C.}\ \bibnamefont {Guo}},\ and\ \bibinfo {author}
  {\bibfnamefont {Z.-F.}\ \bibnamefont {Han}},\ }\bibfield  {title} {\bibinfo
  {title} {Device- and semi--device-independent random numbers based on
  noninequality paradox},\ }\href {https://doi.org/10.1103/PhysRevA.92.022327}
  {\bibfield  {journal} {\bibinfo  {journal} {Phys. Rev. A}\ }\textbf {\bibinfo
  {volume} {92}},\ \bibinfo {pages} {022327} (\bibinfo {year}
  {2015})}\BibitemShut {NoStop}%
\bibitem [{\citenamefont {Rahaman}\ \emph {et~al.}(2015)\citenamefont
  {Rahaman}, \citenamefont {Parker}, \citenamefont {Mironowicz},\ and\
  \citenamefont {Paw\l{}owski}}]{Rahaman2015device}%
  \BibitemOpen
  \bibfield  {author} {\bibinfo {author} {\bibfnamefont {R.}~\bibnamefont
  {Rahaman}}, \bibinfo {author} {\bibfnamefont {M.~G.}\ \bibnamefont {Parker}},
  \bibinfo {author} {\bibfnamefont {P.}~\bibnamefont {Mironowicz}},\ and\
  \bibinfo {author} {\bibfnamefont {M.}~\bibnamefont {Paw\l{}owski}},\
  }\bibfield  {title} {\bibinfo {title} {Device-independent quantum key
  distribution based on measurement inputs},\ }\href
  {https://doi.org/10.1103/PhysRevA.92.062304} {\bibfield  {journal} {\bibinfo
  {journal} {Phys. Rev. A}\ }\textbf {\bibinfo {volume} {92}},\ \bibinfo
  {pages} {062304} (\bibinfo {year} {2015})}\BibitemShut {NoStop}%
\bibitem [{\citenamefont {Ramanathan}\ \emph {et~al.}(2018)\citenamefont
  {Ramanathan}, \citenamefont {Horodecki}, \citenamefont {Anwer}, \citenamefont
  {Pironio}, \citenamefont {Horodecki}, \citenamefont {Gr{\"u}nfeld},
  \citenamefont {Muhammad}, \citenamefont {Bourennane},\ and\ \citenamefont
  {Horodecki}}]{ramanathan2018practical}%
  \BibitemOpen
  \bibfield  {author} {\bibinfo {author} {\bibfnamefont {R.}~\bibnamefont
  {Ramanathan}}, \bibinfo {author} {\bibfnamefont {M.}~\bibnamefont
  {Horodecki}}, \bibinfo {author} {\bibfnamefont {H.}~\bibnamefont {Anwer}},
  \bibinfo {author} {\bibfnamefont {S.}~\bibnamefont {Pironio}}, \bibinfo
  {author} {\bibfnamefont {K.}~\bibnamefont {Horodecki}}, \bibinfo {author}
  {\bibfnamefont {M.}~\bibnamefont {Gr{\"u}nfeld}}, \bibinfo {author}
  {\bibfnamefont {S.}~\bibnamefont {Muhammad}}, \bibinfo {author}
  {\bibfnamefont {M.}~\bibnamefont {Bourennane}},\ and\ \bibinfo {author}
  {\bibfnamefont {P.}~\bibnamefont {Horodecki}},\ }\bibfield  {title} {\bibinfo
  {title} {{Practical No-Signalling proof Randomness Amplification using Hardy
  paradoxes and its experimental implementation}},\ }\href@noop {} {\bibfield
  {journal} {\bibinfo  {journal} {arXiv preprint arXiv:1810.11648}\ } (\bibinfo
  {year} {2018})}\BibitemShut {NoStop}%
\bibitem [{\citenamefont {Rai}\ \emph {et~al.}(2022)\citenamefont {Rai},
  \citenamefont {Pivoluska}, \citenamefont {Sasmal}, \citenamefont {Banik},
  \citenamefont {Ghosh},\ and\ \citenamefont {Plesch}}]{Rai2022self}%
  \BibitemOpen
  \bibfield  {author} {\bibinfo {author} {\bibfnamefont {A.}~\bibnamefont
  {Rai}}, \bibinfo {author} {\bibfnamefont {M.}~\bibnamefont {Pivoluska}},
  \bibinfo {author} {\bibfnamefont {S.}~\bibnamefont {Sasmal}}, \bibinfo
  {author} {\bibfnamefont {M.}~\bibnamefont {Banik}}, \bibinfo {author}
  {\bibfnamefont {S.}~\bibnamefont {Ghosh}},\ and\ \bibinfo {author}
  {\bibfnamefont {M.}~\bibnamefont {Plesch}},\ }\bibfield  {title} {\bibinfo
  {title} {Self-testing quantum states via nonmaximal violation in {Hardy's}
  test of nonlocality},\ }\href {https://doi.org/10.1103/PhysRevA.105.052227}
  {\bibfield  {journal} {\bibinfo  {journal} {Phys. Rev. A}\ }\textbf {\bibinfo
  {volume} {105}},\ \bibinfo {pages} {052227} (\bibinfo {year}
  {2022})}\BibitemShut {NoStop}%
\bibitem [{\citenamefont {Zhao}\ \emph {et~al.}(2023)\citenamefont {Zhao},
  \citenamefont {Ramanathan}, \citenamefont {Liu},\ and\ \citenamefont
  {Horodecki}}]{zhao2023tilted}%
  \BibitemOpen
  \bibfield  {author} {\bibinfo {author} {\bibfnamefont {S.}~\bibnamefont
  {Zhao}}, \bibinfo {author} {\bibfnamefont {R.}~\bibnamefont {Ramanathan}},
  \bibinfo {author} {\bibfnamefont {Y.}~\bibnamefont {Liu}},\ and\ \bibinfo
  {author} {\bibfnamefont {P.}~\bibnamefont {Horodecki}},\ }\bibfield  {title}
  {\bibinfo {title} {Tilted hardy paradoxes for device-independent randomness
  extraction},\ }\href
  {https://doi.org/https://doi.org/10.22331/q-2023-09-15-1114} {\bibfield
  {journal} {\bibinfo  {journal} {Quantum}\ }\textbf {\bibinfo {volume} {7}},\
  \bibinfo {pages} {1114} (\bibinfo {year} {2023})}\BibitemShut {NoStop}%
\bibitem [{\citenamefont {Boschi}\ \emph {et~al.}(1997)\citenamefont {Boschi},
  \citenamefont {Branca}, \citenamefont {De~Martini},\ and\ \citenamefont
  {Hardy}}]{BoschiLadder1997}%
  \BibitemOpen
  \bibfield  {author} {\bibinfo {author} {\bibfnamefont {D.}~\bibnamefont
  {Boschi}}, \bibinfo {author} {\bibfnamefont {S.}~\bibnamefont {Branca}},
  \bibinfo {author} {\bibfnamefont {F.}~\bibnamefont {De~Martini}},\ and\
  \bibinfo {author} {\bibfnamefont {L.}~\bibnamefont {Hardy}},\ }\bibfield
  {title} {\bibinfo {title} {Ladder proof of nonlocality without inequalities:
  Theoretical and experimental results},\ }\href
  {https://doi.org/10.1103/PhysRevLett.79.2755} {\bibfield  {journal} {\bibinfo
   {journal} {Phys. Rev. Lett.}\ }\textbf {\bibinfo {volume} {79}},\ \bibinfo
  {pages} {2755} (\bibinfo {year} {1997})}\BibitemShut {NoStop}%
\bibitem [{\citenamefont {Barbieri}\ \emph {et~al.}(2005)\citenamefont
  {Barbieri}, \citenamefont {{De Martini}}, \citenamefont {{Di Nepi}},\ and\
  \citenamefont {Mataloni}}]{BARBIERI200523towards}%
  \BibitemOpen
  \bibfield  {author} {\bibinfo {author} {\bibfnamefont {M.}~\bibnamefont
  {Barbieri}}, \bibinfo {author} {\bibfnamefont {F.}~\bibnamefont {{De
  Martini}}}, \bibinfo {author} {\bibfnamefont {G.}~\bibnamefont {{Di Nepi}}},\
  and\ \bibinfo {author} {\bibfnamefont {P.}~\bibnamefont {Mataloni}},\
  }\bibfield  {title} {\bibinfo {title} {Towards a test of non-locality without
  “supplementary assumptions”},\ }\href
  {https://doi.org/https://doi.org/10.1016/j.physleta.2004.10.076} {\bibfield
  {journal} {\bibinfo  {journal} {Phys. Lett. A}\ }\textbf {\bibinfo {volume}
  {334}},\ \bibinfo {pages} {23} (\bibinfo {year} {2005})}\BibitemShut
  {NoStop}%
\bibitem [{\citenamefont {Vallone}\ \emph {et~al.}(2011)\citenamefont
  {Vallone}, \citenamefont {Gianani}, \citenamefont {Inostroza}, \citenamefont
  {Saavedra}, \citenamefont {Lima}, \citenamefont {Cabello},\ and\
  \citenamefont {Mataloni}}]{Vallone2011Testing}%
  \BibitemOpen
  \bibfield  {author} {\bibinfo {author} {\bibfnamefont {G.}~\bibnamefont
  {Vallone}}, \bibinfo {author} {\bibfnamefont {I.}~\bibnamefont {Gianani}},
  \bibinfo {author} {\bibfnamefont {E.~B.}\ \bibnamefont {Inostroza}}, \bibinfo
  {author} {\bibfnamefont {C.}~\bibnamefont {Saavedra}}, \bibinfo {author}
  {\bibfnamefont {G.}~\bibnamefont {Lima}}, \bibinfo {author} {\bibfnamefont
  {A.}~\bibnamefont {Cabello}},\ and\ \bibinfo {author} {\bibfnamefont
  {P.}~\bibnamefont {Mataloni}},\ }\bibfield  {title} {\bibinfo {title}
  {Testing {Hardy's} nonlocality proof with genuine energy-time entanglement},\
  }\href {https://doi.org/10.1103/PhysRevA.83.042105} {\bibfield  {journal}
  {\bibinfo  {journal} {Phys. Rev. A}\ }\textbf {\bibinfo {volume} {83}},\
  \bibinfo {pages} {042105} (\bibinfo {year} {2011})}\BibitemShut {NoStop}%
\bibitem [{\citenamefont {Chen}\ and\ \citenamefont
  {Romero}(2012)}]{Chen2012Quantum}%
  \BibitemOpen
  \bibfield  {author} {\bibinfo {author} {\bibfnamefont {L.}~\bibnamefont
  {Chen}}\ and\ \bibinfo {author} {\bibfnamefont {J.}~\bibnamefont {Romero}},\
  }\bibfield  {title} {\bibinfo {title} {Hardy's nonlocality proof using
  twisted photons},\ }\href {https://doi.org/10.1364/OE.20.021687} {\bibfield
  {journal} {\bibinfo  {journal} {Opt. Express}\ }\textbf {\bibinfo {volume}
  {20}},\ \bibinfo {pages} {21687} (\bibinfo {year} {2012})}\BibitemShut
  {NoStop}%
\bibitem [{\citenamefont {Karimi}\ \emph {et~al.}(2014)\citenamefont {Karimi},
  \citenamefont {Cardano}, \citenamefont {Maffei}, \citenamefont {de~Lisio},
  \citenamefont {Marrucci}, \citenamefont {Boyd},\ and\ \citenamefont
  {Santamato}}]{Karimi2014Hardy}%
  \BibitemOpen
  \bibfield  {author} {\bibinfo {author} {\bibfnamefont {E.}~\bibnamefont
  {Karimi}}, \bibinfo {author} {\bibfnamefont {F.}~\bibnamefont {Cardano}},
  \bibinfo {author} {\bibfnamefont {M.}~\bibnamefont {Maffei}}, \bibinfo
  {author} {\bibfnamefont {C.}~\bibnamefont {de~Lisio}}, \bibinfo {author}
  {\bibfnamefont {L.}~\bibnamefont {Marrucci}}, \bibinfo {author}
  {\bibfnamefont {R.~W.}\ \bibnamefont {Boyd}},\ and\ \bibinfo {author}
  {\bibfnamefont {E.}~\bibnamefont {Santamato}},\ }\bibfield  {title} {\bibinfo
  {title} {Hardy's paradox tested in the spin-orbit hilbert space of single
  photons},\ }\href {https://doi.org/10.1103/PhysRevA.89.032122} {\bibfield
  {journal} {\bibinfo  {journal} {Phys. Rev. A}\ }\textbf {\bibinfo {volume}
  {89}},\ \bibinfo {pages} {032122} (\bibinfo {year} {2014})}\BibitemShut
  {NoStop}%
\bibitem [{\citenamefont {Chen}\ \emph {et~al.}(2017)\citenamefont {Chen},
  \citenamefont {Zhang}, \citenamefont {Wu}, \citenamefont {Wang},
  \citenamefont {Fickler},\ and\ \citenamefont
  {Karimi}}]{Chen2017Experimental}%
  \BibitemOpen
  \bibfield  {author} {\bibinfo {author} {\bibfnamefont {L.}~\bibnamefont
  {Chen}}, \bibinfo {author} {\bibfnamefont {W.}~\bibnamefont {Zhang}},
  \bibinfo {author} {\bibfnamefont {Z.}~\bibnamefont {Wu}}, \bibinfo {author}
  {\bibfnamefont {J.}~\bibnamefont {Wang}}, \bibinfo {author} {\bibfnamefont
  {R.}~\bibnamefont {Fickler}},\ and\ \bibinfo {author} {\bibfnamefont
  {E.}~\bibnamefont {Karimi}},\ }\bibfield  {title} {\bibinfo {title}
  {Experimental ladder proof of {Hardy's} nonlocality for high-dimensional
  quantum systems},\ }\href {https://doi.org/10.1103/PhysRevA.96.022115}
  {\bibfield  {journal} {\bibinfo  {journal} {Phys. Rev. A}\ }\textbf {\bibinfo
  {volume} {96}},\ \bibinfo {pages} {022115} (\bibinfo {year}
  {2017})}\BibitemShut {NoStop}%
\bibitem [{\citenamefont {Yang}\ \emph {et~al.}(2019)\citenamefont {Yang},
  \citenamefont {Meng}, \citenamefont {Zhou}, \citenamefont {Xu}, \citenamefont
  {Xiao}, \citenamefont {Sun}, \citenamefont {Chen}, \citenamefont {Xu},
  \citenamefont {Li},\ and\ \citenamefont {Guo}}]{Yang2019Stronger}%
  \BibitemOpen
  \bibfield  {author} {\bibinfo {author} {\bibfnamefont {M.}~\bibnamefont
  {Yang}}, \bibinfo {author} {\bibfnamefont {H.-X.}\ \bibnamefont {Meng}},
  \bibinfo {author} {\bibfnamefont {J.}~\bibnamefont {Zhou}}, \bibinfo {author}
  {\bibfnamefont {Z.-P.}\ \bibnamefont {Xu}}, \bibinfo {author} {\bibfnamefont
  {Y.}~\bibnamefont {Xiao}}, \bibinfo {author} {\bibfnamefont {K.}~\bibnamefont
  {Sun}}, \bibinfo {author} {\bibfnamefont {J.-L.}\ \bibnamefont {Chen}},
  \bibinfo {author} {\bibfnamefont {J.-S.}\ \bibnamefont {Xu}}, \bibinfo
  {author} {\bibfnamefont {C.-F.}\ \bibnamefont {Li}},\ and\ \bibinfo {author}
  {\bibfnamefont {G.-C.}\ \bibnamefont {Guo}},\ }\bibfield  {title} {\bibinfo
  {title} {{Stronger Hardy-type paradox based on the Bell inequality and its
  experimental test}},\ }\href {https://doi.org/10.1103/PhysRevA.99.032103}
  {\bibfield  {journal} {\bibinfo  {journal} {Phys. Rev. A}\ }\textbf {\bibinfo
  {volume} {99}},\ \bibinfo {pages} {032103} (\bibinfo {year}
  {2019})}\BibitemShut {NoStop}%
\bibitem [{\citenamefont {Das}\ and\ \citenamefont {Paul}(2020)}]{das2020new}%
  \BibitemOpen
  \bibfield  {author} {\bibinfo {author} {\bibfnamefont {S.}~\bibnamefont
  {Das}}\ and\ \bibinfo {author} {\bibfnamefont {G.}~\bibnamefont {Paul}},\
  }\bibfield  {title} {\bibinfo {title} {A new error-modeling of {Hardy’s
  Paradox} for superconducting qubits and its experimental verification},\
  }\href {https://dl.acm.org/doi/abs/10.1145/3396239} {\bibfield  {journal}
  {\bibinfo  {journal} {ACM Trans. Quantum Comput.}\ }\textbf {\bibinfo
  {volume} {1}},\ \bibinfo {pages} {1} (\bibinfo {year} {2020})}\BibitemShut
  {NoStop}%
\bibitem [{\citenamefont {Zhang}\ \emph {et~al.}(2011)\citenamefont {Zhang},
  \citenamefont {Glancy},\ and\ \citenamefont
  {Knill}}]{Zhang2011Asymptotically}%
  \BibitemOpen
  \bibfield  {author} {\bibinfo {author} {\bibfnamefont {Y.}~\bibnamefont
  {Zhang}}, \bibinfo {author} {\bibfnamefont {S.}~\bibnamefont {Glancy}},\ and\
  \bibinfo {author} {\bibfnamefont {E.}~\bibnamefont {Knill}},\ }\bibfield
  {title} {\bibinfo {title} {Asymptotically optimal data analysis for rejecting
  local realism},\ }\href {https://doi.org/10.1103/PhysRevA.84.062118}
  {\bibfield  {journal} {\bibinfo  {journal} {Phys. Rev. A}\ }\textbf {\bibinfo
  {volume} {84}},\ \bibinfo {pages} {062118} (\bibinfo {year}
  {2011})}\BibitemShut {NoStop}%
\bibitem [{\citenamefont {Eberhard}(1993)}]{Eberhard1993Background}%
  \BibitemOpen
  \bibfield  {author} {\bibinfo {author} {\bibfnamefont {P.~H.}\ \bibnamefont
  {Eberhard}},\ }\bibfield  {title} {\bibinfo {title} {Background level and
  counter efficiencies required for a loophole-free {Einstein-Podolsky-Rosen}
  experiment},\ }\href {https://doi.org/10.1103/PhysRevA.47.R747} {\bibfield
  {journal} {\bibinfo  {journal} {Phys. Rev. A}\ }\textbf {\bibinfo {volume}
  {47}},\ \bibinfo {pages} {R747} (\bibinfo {year} {1993})}\BibitemShut
  {NoStop}%
\bibitem [{\citenamefont {Hwang}\ \emph {et~al.}(1996)\citenamefont {Hwang},
  \citenamefont {Koh},\ and\ \citenamefont {Han}}]{HWANG1996The}%
  \BibitemOpen
  \bibfield  {author} {\bibinfo {author} {\bibfnamefont {W.~Y.}\ \bibnamefont
  {Hwang}}, \bibinfo {author} {\bibfnamefont {I.~G.}\ \bibnamefont {Koh}},\
  and\ \bibinfo {author} {\bibfnamefont {Y.~D.}\ \bibnamefont {Han}},\
  }\bibfield  {title} {\bibinfo {title} {The detection loophole in hardy's
  nonlocality theorem and minimum detection efficiency},\ }\href
  {https://doi.org/https://doi.org/10.1016/0375-9601(96)00072-2} {\bibfield
  {journal} {\bibinfo  {journal} {Phys. Lett. A}\ }\textbf {\bibinfo {volume}
  {212}},\ \bibinfo {pages} {309} (\bibinfo {year} {1996})}\BibitemShut
  {NoStop}%
\bibitem [{\citenamefont {Ghirardi}\ and\ \citenamefont
  {Marinatto}(2006)}]{Ghirardi2006Hardy}%
  \BibitemOpen
  \bibfield  {author} {\bibinfo {author} {\bibfnamefont {G.}~\bibnamefont
  {Ghirardi}}\ and\ \bibinfo {author} {\bibfnamefont {L.}~\bibnamefont
  {Marinatto}},\ }\bibfield  {title} {\bibinfo {title} {Hardy's proof of
  nonlocality in the presence of noise},\ }\href
  {https://doi.org/10.1103/PhysRevA.74.062107} {\bibfield  {journal} {\bibinfo
  {journal} {Phys. Rev. A}\ }\textbf {\bibinfo {volume} {74}},\ \bibinfo
  {pages} {062107} (\bibinfo {year} {2006})}\BibitemShut {NoStop}%
\bibitem [{\citenamefont {Braun}\ and\ \citenamefont
  {Choi}(2008)}]{Braun2008Hardy}%
  \BibitemOpen
  \bibfield  {author} {\bibinfo {author} {\bibfnamefont {D.}~\bibnamefont
  {Braun}}\ and\ \bibinfo {author} {\bibfnamefont {M.-S.}\ \bibnamefont
  {Choi}},\ }\bibfield  {title} {\bibinfo {title} {{Hardy's test versus the
  Clauser-Horne-Shimony-Holt test of quantum nonlocality: Fundamental and
  practical aspects}},\ }\href {https://doi.org/10.1103/PhysRevA.78.032114}
  {\bibfield  {journal} {\bibinfo  {journal} {Phys. Rev. A}\ }\textbf {\bibinfo
  {volume} {78}},\ \bibinfo {pages} {032114} (\bibinfo {year}
  {2008})}\BibitemShut {NoStop}%
\bibitem [{\citenamefont {Kullback}\ and\ \citenamefont
  {Leibler}(1951)}]{kullback1951information}%
  \BibitemOpen
  \bibfield  {author} {\bibinfo {author} {\bibfnamefont {S.}~\bibnamefont
  {Kullback}}\ and\ \bibinfo {author} {\bibfnamefont {R.~A.}\ \bibnamefont
  {Leibler}},\ }\bibfield  {title} {\bibinfo {title} {On information and
  sufficiency},\ }\href {https://www.jstor.org/stable/2236703} {\bibfield
  {journal} {\bibinfo  {journal} {Ann. Math. Stat.}\ }\textbf {\bibinfo
  {volume} {22}},\ \bibinfo {pages} {79} (\bibinfo {year} {1951})}\BibitemShut
  {NoStop}%
\end{thebibliography}%
%%%%%%%%%%%%%%%%%%%%%%%%%%%%%%%%%%%%%%%%

%\input{.bbl}
\end{document}